\documentclass{ws1-procs9x6}  

\usepackage{epsfig}

\newlength{\sizeonefig}
\newlength{\sizetwofig}
\begin{document}

\newcommand{\ket}[1]{{#1}}
\newcommand{\braket}[2]{{\langle #1,#2 \rangle}}

\newcommand{\myremark}[1]{\textbf{#1}}

\newcommand{\beq}{\begin{equation}}
\newcommand{\eeq}{\end{equation}}
\newcommand{\bea}{\begin{eqnarray}} 
\newcommand{\eea}{\end{eqnarray}}

\def\laq{\raise 0.4ex\hbox{$<$}\kern -0.8em\lower 0.62ex\hbox{$\sim$}}
\def\gaq{\raise 0.4ex\hbox{$>$}\kern -0.7em\lower 0.62ex\hbox{$\sim$}}

\def\tdot#1{{\buildrel{\ldots}\over{#1}}}

\title{TASI Lectures on Gravitational Waves from the Early Universe} 

\author{Alessandra Buonanno}

\address{groupe de Gravitation et Cosmologie (GReCO)\\
Institut d'Astrophysique de Paris (CNRS)\\  
98$^{\rm bis}$ Boulevard Arago, 75014 Paris, France\\
\small email: {\tt buonanno@iap.fr}}

\maketitle

\abstracts{These lectures discuss how the direct detection 
of gravitational waves can be used to probe the very 
early Universe. We review the main cosmological mechanisms 
which could have produced relic gravitational waves, 
and compare theoretical predictions with  
capabilities and time scales of current and 
upcoming experiments.}

\section{Overview of gravitational-wave research}
\label{sec1}

In 1916 Einstein realized the propagation effects at finite 
velocity in the gravitational equations and predicted the 
existence of wave-like solutions of the linearized vacuum 
field equations~\refcite{E16}. The work of Bondi~\refcite{HB} 
in the mid 50s, applied to self-gravitating systems like 
binaries made of neutron stars and/or black holes, 
proved that gravitational waves (GWs) carry off energy from those systems,  
as the 1974 discovery of the binary pulsar PSR 1913+16 by 
Hulse and Taylor~\refcite{HT75} confirmed indisputably.  
This discovery is an {\it indirect} observation of GWs. In fact, the speed 
up of the two-body orbital period can be explained 
as due to GW emission. 

The experimental search for {\it direct} observation of gravitational waves 
begun only in the 60s with the pioneering work of Joseph Weber, 
and for almost three decades it has been pursued solely using meter-scale 
resonant-bar detectors. These detectors are cylindrical-shape  bars 
whose mechanical oscillations can be driven 
by GWs. Five cryogenic resonant-bar detectors are in operation since 1990: 
ALLEGRO, AURIGA, EXPLORER, NAUTILUS and NIOBE~\refcite{bars}. They are 
narrow-band detectors sensitive to GW frequency $\sim$kHz. During the last years a 
world-wide network of kilometer-scale 
ground-based laser-interferometer detectors has been built and 
has begun operation in Japan (TAMA\,300), in the United States (LIGOs) 
and Europe (GEO\,600 and VIRGO)~\refcite{IFOs,HMBH01}. The frequency band is 
$\sim 1\mbox{--} 10^3$ Hz. Meantime the first 
detectors begin the search, development of the next generation 
kilometer-scale interferometers is already underway
~\refcite{AdIFO2s,ligoII,AdIFO3s,ligoIII}. 
A laser-interferometer space antenna (LISA)~\refcite{LISA,HMBH01} is also planned 
by the European Space Agency (ESA) and NASA, though not yet fully funded, 
and could fly in $~\sim 2011$. It consists of three drag-free spacecrafts  
in a triangle configuration. The spacecraft tracks the distance 
to each other ($\sim$ 5 million kms) using laser beams, searching 
GWs in the frequency range $\sim 10^{-4} \mbox{--} 10^{-1}$ Hz.   
Thanks to the theoretical and experimental progress made 
by the GW community during the last forty years, the direct detection of 
GWs is not far ahead of us and, hopefully, it will mark the next decade.

The most promising sources for both ground- and space-based detectors 
are astrophysical sources at relatively low red-shift~\refcite{CT02}, like binaries made 
of black holes (BHs) and/or neutron stars (NSs), white-dwarf (WD) binaries, supermassive 
black holes, low-mass X-ray binaries, supernovae and 
pulsars. The large scale interferometers were indeed conceived and designed 
to detect GWs from those sources. These lectures will focus on 
cosmological sources at much higher red-shift $z \gg 1$. 
We shall discuss if and how the detection of GWs can be used to 
investigate physical processes occured around Universe's birth. 
Various excellent lectures and reviews were written on this subject like, 
e.g., the ones by Allen~\refcite{A96} and Maggiore~\refcite{M00} 
[see also Sections 2.9 and 3.6 by Cutler and Thorne~\refcite{CT02}]. 

It is not possible to cover in two lectures 
all the literature on the subject, so we have to make a selection. 
In the first lecture, after briefly reviewing the key ideas  
underlying GW detectors [see Section~\ref{sec2}], we discuss general 
features of relic GW spectra, their typical intensity and range of 
frequencies and phenomenological bounds [see Section~\ref{sec3}]. The bulk 
of the lecture will be the analysis of the phenomenon of amplification 
of quantum-vacuum fluctuations, and the evaluation of stochastic GW backgrounds 
in standard, quintessential and superstring-motivated inflationary scenarios 
[see Section~\ref{sec4}]. The second lecture will review, in Section~\ref{sec5}, 
other physical mechanisms which could produce GWs, like first-order 
phase transitions, turbulence, 
preheating, etc., and cosmic strings [see Section~\ref{sec6}]. Then, 
in Section~\ref{sec7} we discuss which differences in the relic GW spectrum we could expect 
from brane-world scenarios and, finally, in Section~\ref{sec8} we comment  
on the possibility of extracting the red-shift--luminosity curve and 
cosmological parameters by detecting GWs from binary systems.
Section~\ref{sec9} summarizes the conclusions. 

\section{Key ideas underlying GW detectors}
\label{sec2}

The simplest GW detector we can imagine is a body of mass $m$ at a 
distance $L$ from a fiducial laboratory point, connected to it 
by a spring of resonant frequency $\omega_0$ and quality 
factor $Q$. Einstein equation of geodesic deviation predicts that 
the infinitesimal displacement $\Delta L$ of the mass  along the 
line of separation from its equilibrium position satisfies the 
equation~\refcite{KT87} (valid for GW wavelengths $\gg L$ and 
in local Lorentz frame of the observer at the fiducial 
laboratory point) 
\beq
\ddot{\Delta L}(t) + 2\frac{\omega_0}{Q}\,\dot{\Delta L}(t) + \omega_0^2\,\Delta L(t) = 
\frac{L}{2}\,\left [F_+\,\ddot{h}_+(t) + F_\times\,\ddot{h}_\times(t) \right ]\,,
\label{sd}
\eeq
where $F_{+,\times}$ are coefficients of order unity 
which depend on the direction of the source and the 
GW polarization angle; $h_{+,\times}$ are the two independent 
polarizations of the GW. 

Laser-interferometer GW detectors, such as GEO, LIGO, VIRGO and TAMA, 
are composed of two perpendicular kilometer-scale arm cavities with two 
test-mass mirrors hung by wires at the end of each cavity.
The tiny displacements $\Delta L$ of the mirrors induced by a passing-by GW 
are monitored with very high accuracy by measuring the relative optical phase 
between the light paths in each interferometer arm.
The mirrors are pendula with quality factor $Q$ quite high and 
resonant frequency $\omega_0$ much lower ($\sim 1$ Hz) than the typical 
GW frequency ($\sim 100$ Hz). In this case Eq.~(\ref{sd}), written 
in Fourier domain, reduces to $\Delta L/L \sim h$. 
The typical amplitude, at $100$ Hz, of GWs emitted by binary systems 
in the VIRGO cluster of galaxies (15 Mpc distant), 
which is the largest distance the first-generation of 
ground-based interferometers can probe, is $\sim 10^{-21}$. 
This means $\Delta L \sim 10^{-18}$ m, a very tiny number! 
It seems rather discouraging, especially if we think to monitor 
the test-mass motion with light of wavelength nearly $10^{12}$ times 
larger. It took a long time, theoretically and experimentally 
to develop sophisticated technology to achieve measurements of 
such tiny displacements. 

The electromagnetic  signal coming out the interferometer will contain 
the GW signal but also noise -- 
for example thermal noise in the suspension system and 
in the mirror itself, can shake the mirror mimicking the effect 
of a GW. The root-mean-square of the noise is generally  expressed in terms 
of the noise power per unit frequency $S_n$ by the relation 
$h_{\rm rms} \sim \sqrt{S_n(f)\,\Delta f} \sim \Delta L/L$, $\Delta L$ being the 
mirror displacement induced by noise and $\Delta f$ the frequency bandwidth. 
 
\section{Production of relic gravitational waves: general features}
\label{sec3}

\subsection{Gravitational-wave spectrum}
\label{sec3.1}

Since in the following we will focus mainly on the primordial 
stochastic background of GWs, in this section we relate its energy density to the  
{\it spectral density} $S_h$, which is the quantity of direct interest when we 
want to compare theoretical predictions to experimental sensitivities. 
[Henceforth, we pose $\hbar=1=c$.]

The intensity of a stochastic background of GWs at (present) 
frequency $f$ is generally expressed in terms of 
\beq
\label{intGW}
\Omega_{\rm GW}(f) = \frac{1}{\rho_c}\,\frac{d \rho_{\rm GW}}{d \log f}\,,
\eeq
where $\rho_c=3H_0^2/(8\pi G_{\rm N})$ and $\rho_{\rm GW}$ is 
the energy density of the GW stochastic background.~\footnote{
The value of $H_0$ is generally written as $H_0=h_0\times 100$ 
km/sec/Mpc where $h_0$ parameterizes the existing experimental 
uncertainty. Henceforth, all primordial 
spectrum will be expressed in terms of $h_0^2\,\Omega_{\rm GW}$.}  
In the transverse-traceless (TT) gauge, denoting with $\hat{\Omega}$ 
the direction of propagation of the GW, with $\hat{m}, \hat{n}$ the 
unit vectors forming with $\hat{\Omega}$ an orthonormal basis, 
and with $\epsilon_{i j}^+=\hat{m}_i\,\hat{m}_j-
\hat{n}_i\,\hat{n}_j, \epsilon_{i j}^\times=\hat{m}_i\,\hat{n}_j+
\hat{n}_i\,\hat{m}_j$ the polarization tensors, $i,j=1,2,3$  
[$\epsilon_{i j}^{\rm P}(\Omega)\,e^{\rm Q\,ij}(\Omega) = 2 \delta^{\rm PQ}$],  
a stochastic, isotropic GW can be written as
\beq
h_{i j}(t) = \sum_{\rm P=+,\times}\,\int_{-\infty}^{+\infty} 
df\,\int d \hat{\Omega}\,h_{\rm P}(f,\hat{\Omega})\, 
e^{-2\pi i f t}\,\epsilon_{i j}^{\rm P}(\hat{\Omega})\,,
\label{hGW}
\eeq
where $d\hat{\Omega} = d\phi\,d\cos \theta$, 
$h_{\rm P}(f,\hat{\Omega}) = h_{\rm P}^{\,*}(-f,\hat{\Omega})$, and 
in writing Eq.~(\ref{hGW}) we have assumed, in full generality, 
that the wave is located at $\vec{x}=0$. 
If the stochastic background is also stationary and unpolarized, 
averaging the Fourier components over an ensemble gives~\refcite{M00}
\beq
<h_{\rm P}^*(f,\hat{\Omega})\,h_{\rm Q}(f^\prime,\hat{\Omega}^\prime)>= 
\delta(f-f^\prime)\,\delta^2(\hat{\Omega},\hat{\Omega}^\prime)\,\frac{1}{4 \pi}\,
\delta_{\rm PQ}\,\frac{1}{2}\,S_h(f)\,,
\label{aver}
\eeq
with $\delta^2(\hat{\Omega},\hat{\Omega}^\prime)=\delta (\cos \theta 
-\cos \theta^\prime)\,\delta(\phi-\phi^\prime)$. 
The above equation defines the (one-sided) {\it spectral density} $S_h$ 
[$S_h(f)=S_h(-f)$], which has the dimensions of Hz$^{-1}$. 
The energy density of GWs is given by the 00 component of the 
energy-momentum tensor averaged over several wavelengths [see 
e.g., Ref.~\refcite{KT87}]
\begin{figure}[t]
\begin{center}
\includegraphics[width=0.7\textwidth]{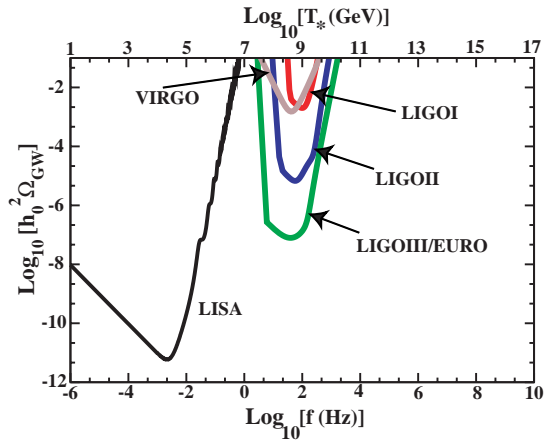}
\vspace{-0.3cm}
\caption{\label{ranges} Sensitivities, expressed 
in terms of $\Omega_{\rm GW}$, versus frequency of LISA 
and ground-based detectors of first, second and third generation.
On the top axis, the estimated 
temperature of the Universe when GWs are produced by causal 
mechanisms during RD era [see Eq.~(\protect\ref{fRD})].}
\end{center}
\end{figure}
\beq
\rho_{\rm GW} = \int_0^{\infty} d \log f\, \frac{d \rho_{\rm GW}}{d \log f}=
\frac{1}{32 \pi\,G_{\rm N}}\,\overline{\dot{h_{ij}}\,\dot{h}^{i j}}\,.
\label{rhoGW}
\eeq
Inserting in the above equation Eq.~(\ref{hGW}) and using the property that 
for a stochastic background the spatial average can be traded with the ensemble 
average defined by Eq.~(\ref{aver}), a straightforward calculation leads to: 
\beq
\label{OmSh}
\Omega_{\rm GW}(f) = \frac{4 \pi^2}{3 H_0^2}\,f^3\,S_h(f)\simeq
1.25 \times 10^{36}\,\frac{1}{h_0^2}\,\left (\frac{f}{\rm Hz} \right)^2\,f\,S_h(f)\,.
\eeq
Note that if $\Omega_{\rm GW} \sim \mathcal{C}\,f^\alpha$ with 
$0 \leq \alpha < 3$, due to the factor $f^3$ appearing in the RHS of 
Eq.~(\ref{OmSh}), a GW detector operating at lower frequency needs lower sensitivity 
than a GW detector operating at higher frequency. This explains why space-based 
detectors like LISA can have better performances than ground-based detectors, 
although the expected sensitivity of LISA is worse than LIGO and VIRGO.

In Fig.~\ref{ranges} we plot the sensitivities for ground-based 
detectors of first~\refcite{IFOs}, second ($\sim 2008$)~\refcite{AdIFO2s,ligoII} and 
third ($> 2011$)~\refcite{AdIFO3s,ligoIII} 
generation. More specifically, for the second generation we 
use the Advanced LIGO configuration or LIGOII 
~\refcite{AdIFO2s,ligoII}, for which 
considerable research and development (R\&D) have been gained  
during the last years, and for the third generation we 
use, as an example, the speed-meter configuration~\refcite{AdIFO3s,ligoIII}. 
The R\&D for third generation of interferometers as LIGOIII/EURO started 
only more recently.

A stochastic background is a random process which is intrinsically 
indistinguishable from the detector noise. As we shall see in the 
following, the GW signal is expected 
to be far too low to exceed the noise level in any existing or planned 
single detector on the earth. Moreover, the instrumental noise level 
will not be known sufficiently well a priori to search 
for excess noise in each instrument. Therefore, the optimal 
strategy which has been proposed~\refcite{GWdet} is to perform 
a correlation between two or more detectors, possibly widely 
separated to minimize common noise sources. By correlating 
two detectors the increase in sensitivity $h_{\rm rms}$ is 
$\sim (\Delta f\,T)^{1/4}$ where $\Delta f$ is the bandwidth 
and $T$ the total observation time. Henceforth, we shall show the 
sensitivities obtained correlating two ground-based detectors. 
Assuming a constant GW spectrum, and correlating for fourth months, 
we obtain at $90\%$ confident level for two LIGOI: 
$h_0^2\,\Omega_{\rm GW} \simeq 3.5 \times 10^{-6}$, 
for two LIGOII: $h_0^2\,\Omega_{\rm GW} \simeq 5.1 \times 10^{-9}$ and  
for two LIGOIII: $h_0^2\,\Omega_{\rm GW} \simeq 3.7 \times 10^{-11}$. 

In the case of LISA, since only one space-based detector 
is currently planned, the method of correlating two detectors 
cannot be used. [In Ref.~\refcite{CL} optimal orbital alignements
for correlating a pair of future LISA-kind detectors have 
been investigated.] Armstrong, Estabrook and Tinto~\refcite{AET01} realized that  
by combining the signals from the three spacecrafts which compose LISA, 
it is possible to measure the instrumental noise power. Then, 
any excess noise above this power will be due to GWs. However, 
besides the primordial stochastic GW background, there are various  
astrophysically  generated stochastic GW background due to 
binaries present in our galaxy or outside it. These stochastic 
backgrounds are produced when the incoherent superposition of 
gravitational radiation emitted by a large number 
of astrophysical sources cannot be resolved individually.  
A possible, although rather challenging 
way of discriminating between the primordial 
and astrophysically generated backgrounds was suggested by 
Ungarelli and Vecchio~\refcite{UV201}. 
As LISA rotates along its orbit, its sensitivity 
to different directions changes, so LISA could measure 
the isotropy of the stochastic background and use it to separate 
the primordial stochastic background, which is isotropic, 
from the galactic stochastic background, which is anisotropic, 
being concentrated mostly in the galactic plane (rather than in the halo). 
In Fig.~\ref{ranges} we plot the planned LISA sensitivity~\refcite{SP}.

\subsection{When gravitons decoupled: thermal spectrum?}
\label{sec3.2}

It is well known that particles which decoupled from primordial plasma at time 
$t_\mathrm{dec}$, when the Universe had temperature $T_\mathrm{dec}$,
 carry information on the state of the Universe at $T_\mathrm{dec}$.
The weaker the interaction, the higher the energy scale when they drop 
out of thermal equilibrium. To estimate $T_\mathrm{dec}$ for gravitons, 
we proceed as follows~\refcite{ZN,KT90}. We assume that gravitons are in 
thermal equilibrium in the primordial plasma through point-like 
four-body interactions involving two gravitons. 
We denote with $\Gamma=n\,\sigma\,|v|$ the interaction rate per particle, 
$n$ being the number density of particles in equilibrium, $\sigma$ 
the scattering cross section, and $v$ the relative velocity. 
From dimensional considerations we expect $\sigma \sim 
G_\mathrm{N}^2\,\ket{E^2}\sim G_\mathrm{N}^2\,T^2$, 
where $G_\mathrm{N}$ is the Newton constant and $\ket{E^2}$ 
is the average energy squared. For relativistic particles in equilibrium at temperature $T$, 
$n \sim T^3$. We also assume $v \sim 1$. 
The interaction rate is then $\Gamma \sim G_\mathrm{N}^2\,T^5$. 
Assuming the Universe evolves adiabatically, we have $g_\mathrm{S}\,T^3\,a^3= 
{\rm const.}$, where $g_\mathrm{S}$ is the number of effectively 
massless degrees of freedom [see Eq.~(3.73) of Ref.~\refcite{KT90}]. 
Since for relativistic particles $\rho \sim T^4$, we have 
$\dot{T}/{T} \propto H \propto T^2/M_\mathrm{Pl}$, thus 
the condition to maintain thermal equilibrium is 
$\Gamma \gtrsim H$, i.e. the interaction time-scale must 
exceed the local Hubble time. This gives:
\beq
\left ( \frac{\Gamma}{H} \right )_\mathrm{g} \simeq \left ( \frac{T}{M_\mathrm{Pl}}\right )^3.
\eeq 
So, as it could have been anticipated from dimensional considerations, 
gravitons decoupled at $T_{\rm dec} \sim M_\mathrm{Pl}$. Let us derive at which 
temperature it corresponds today. The temperature of the Universe, 
i.e. of the particles in thermal equilibrium, scales as 
$T \propto g_\mathrm{S}^{-1/3}\,a^{-1}$, while the temperature associated 
to particles which have decoupled from the thermal bath 
drops as $T \propto a^{-1}$. 
Applying these considerations to the thermal bath of gravitons, and 
using the fact that at $T \,\gaq\, 300$ GeV the Standard Model (SM) of particle 
physics predicts $g_\mathrm{S} \simeq 106.75$, while at present time, 
assuming three neutrino species, $g_\mathrm{S} \simeq 3.91$, 
it is easily derived that the graviton temperature today is at most
~\refcite{KT90} $T_{\rm g} = (3.91/106.75)^{1/3}\,2.75\,{\rm K} \simeq 0.9$ K 
[at most because the effective number of massless degrees of freedom 
could be higher then $106.75$ at Planckian energies.]. 
Thus, the thermal spectrum of gravitons ranges in the GHz and 
is given by~\refcite{A96} 
\beq
\Omega_g(f) = \frac{8\pi h}{c^3}\,\left (\frac{f}{f_g}\right )^4\,
\frac{1}{\rho_c}\,\frac{f_g^4}{e^{f/f_g}-1} \simeq 2.9 \times 10^{-8}\,
h_{0}^{-2}\,
\left (\frac{f}{f_g}\right )^4\,\frac{e-1}{e^{f/f_g}-1}\,,
\eeq
with $f_g= k\,T_{\rm g}/h\simeq 1.9\times 10^{10}$ Hz. 
[Note that for CMBR we have $T_\gamma \simeq 2.73$ K and 
$f_\gamma = 5.7 \times 10^{10}$ Hz.] 
However, since gravitons interact very weakly with matter 
and radiation, the time required to reach thermal equilibrium 
could be longer than the Hubble time. So, it is quite 
unlikely that this radiation ever existed. Moreover, if 
the Universe underwent a period of inflation after the Planck era, 
the background should have been strongly diluted. 
More importantly, the theory of gravity at 
the Planck energy could differ significantly from Einstein 
theory; if so, the above analysis should be modified, accordingly.

\subsection{Gravitational waves produced by casual mechanisms: 
typical frequencies}
\label{sec3.3}

Two features determine the typical frequency 
of GWs of cosmological origin produced by some casual mechanism~\refcite{M00}: 
(i) the dynamics, which is model dependent, and (ii) the kinematics, 
that is the red-shift from the production era.

Let us assume that a graviton is produced with frequency 
$f_*$ at time $t_*$ during RD or MD era. We want to evaluate 
its frequency today $f = f_*\,a_*/a_0$. Assuming 
that the Universe evolved adiabatically, that is 
$g_\mathrm{S}(T_*)\,T_*^3\,a_*^3 = g_\mathrm{S}(T_0)\,T_0^3\,a_0^3$, 
and using $T_0 = 2.73 K$ and $g_\mathrm{S} = 3.91$ we get
\beq
f \simeq 10^{-13}\,f_*\,\left ( \frac{100}{g_{\mathrm{S}\,*}}\right )^{1/3}\,
\left ( \frac{1 \mathrm{GeV}}{T_*} \right )\,.
\eeq
Since the size of the Hubble radius is the 
length scale beyond which causal microphysics cannot operate, 
from causality considerations we expect the characteristic 
wavelength of gravitons produced at time $t_*$ is either 
$\sim H_*^{-1}$ or smaller. So, we pose~\refcite{M00} 
$\lambda_* = \epsilon\,H_*^{-1}$ with $\epsilon \,\laq\, 1$. 
If the GW signal is produced during RD era $H_*^2 = 
8 \pi G_\mathrm{N}\,\rho_\mathrm{rad}/3=
8\pi^3g_*\,T_*^4\,/(90\,M_\mathrm{Pl}^2)$ and 
\beq
f \simeq 10^{-7}\,\frac{1}{\epsilon}\,
\left ( \frac{T_*}{1\,\mathrm{GeV}}\right )\, 
\left (\frac{g_*}{100}\right )^{1/6}\,\mathrm{Hz}\,.
\label{fRD}
\eeq
On the top axis of Fig.~\ref{ranges} we show the production 
temperature $T_*$ as obtained from Eq.~(\ref{fRD}) by expressing  
$T_*$ in terms of the frequency $f$. For the kind 
of mechanism analysed in this section, LISA could probe 
physics at TeV scale, ground-based interferometers in the range 
$\sim 10^8\mbox{--}10^{10}$ GeV, while GUT and 
Planck scales could be probed in GHz region, where however  
no GW detectors have been planned so far, but electromagnetic (EM) microwave 
cavities have been proposed~\refcite{EM}.

\subsection{Phenomenological bounds}
\label{sec3.4}

\begin{figure}[t]
\begin{center}
\includegraphics[width=0.7\textwidth]{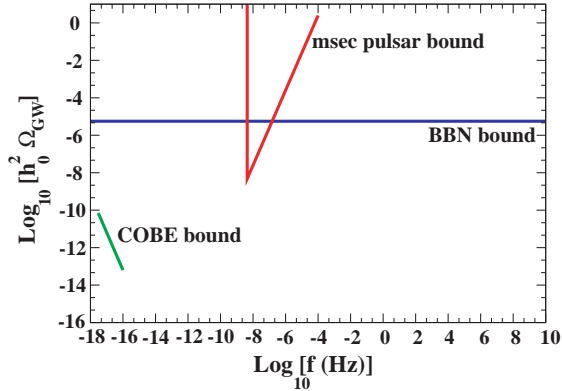}
\caption{\label{fig:bounds} Summary of phenomenological  
bounds on energy density of relic GWs.}
\end{center}
\end{figure}

\subsubsection{Big-bang nucleosynthesis bound}
\label{sec3.4.1}

The theory of Big-bang nucleosynthesis (BBN)~\refcite{KO} predicts rather successfully 
the primordial abundances of light elements like deuterium, 
${}^3$He, ${}^4$He and ${}^7$Li. If at nucleosynthesis time 
($T\simeq $ MeV), the contribution of the primordial GWs to the total energy 
density is too large, then the expansion rate of the Universe $H$, 
and the freeze-out temperature which determines the relative abundance of 
neutrons and protons, will be too high. Thus, neutrons will be 
more available and ${}^4$He will be overproduced, spoiling BBN predictions. 
Detailed calculations provides the following 
bound on the energy density in GWs integrated over frequency, 
\beq
\int_{f=0}^{f=+\infty} d \log f\,h_0^2\,\Omega_{\rm GW}(f) 
\leq 5.6 \times 10^{-6}\,(N_\nu -3)\,,
\label{BBNbound}
\eeq
where $N_\nu$ is the effective number of neutrino species. 
The bound (\ref{BBNbound}) holds only for GWs that were already produced 
at time of nucleosynthesis. Since the integral cannot exceed the bound, 
also its positive definite integrand cannot do it, 
unless $\Omega_{\rm GW} \ll 10^{-6}$ 
for most of the frequency range and has a very narrow peak 
on the order of $\sim 10^{-5}$ at some frequency. 
This possibility is however not very plausible. 
Thus, Eq.~(\ref{BBNbound}) gives for $\Delta f \sim f$ and 
$N_\nu< 4$~\refcite{CST97} the value $h_0^2\,\Omega_{\rm GW} \leq 
5.6 \times 10^{-6}$, which is plotted in Fig.~\ref{fig:bounds}.

\subsubsection{COBE bound}
\label{sec3.4.2}

The COBE bound comes from the measurement of temperature fluctuations 
in the Cosmic Microwave Background Radiation (CMBR).  
Indeed, a background of GWs at very low frequencies 
can produce, if sufficiently strong, a stochastic red-shift 
of the CMBR frequencies, through the well known Sachs-Wolfe 
effect. Indicating by $\delta T$ the fluctuation induced in the 
CMBR temperature, we have
\beq
\Omega_{\rm GW}(f) \leq \left (\frac{H_0}{f}\right )^2\,
\left ( \frac{\delta T}{T}\right )^2\, \quad 3 \times 10^{-18}\,{\rm Hz} 
< f < 10^{-16}\,{\rm Hz}\,,
\eeq
where the range of frequencies is determined by the condition the 
GWs are inside the Hubble radius today ($f > H_0 \sim 3\times 10^{-18}\,{\rm Hz}$) 
and they were outside the Hubble radius at the last scattering surface 
(LSS) ($f< \sqrt{z_{\rm LSS}}\,H_0 \sim 10^{-16}\,{\rm Hz}$ with $z_{\rm LSS} \sim 10^3$).
The value of $\delta T/T$ induced by GWs on CMBR cannot exceed the observed 
value $\delta T/T \sim 5 \times 10^{-6}$. Detailed analyses give~\refcite{AK94,A96}:
\beq
h_0^2\,\Omega_{\rm GW}(f) < 7 \times 10^{-11}\,\left (\frac{H_0}{f}\right )^2 
\quad 3 \times 10^{-18}\,{\rm Hz} < f < 10^{-16}\,{\rm Hz}\,.
\label{cobebound}
\eeq
GWs can saturate this bound if the contribution from scalar 
perturbations is subdominant, and this depends on the specific 
inflationary model. The COBE bound (\ref{cobebound}) is shown 
in Fig.~\ref{fig:bounds}.

\subsubsection{msec pulsar bound}
\label{sec3.5.3}

The very accurate timing of msec pulsars constraints $\Omega_{\rm GW}$.
In fact, if a GW passes between us and the pulsar the time of arrival of the pulse is 
(doppler) shifted. Many years of observations lead to the bound~\refcite{TD96,M00}
\beq
h_0^2\,\Omega_{\rm GW}(f) \leq 4.8 \times 10^{-9}\,\left (\frac{f}{f_*}
\right )^2 \quad \quad f > f_* \equiv 4.4 \times 
10^{-9}\,{\rm Hz}\,,
\eeq
where $f_*$ is derived from the total observation time of 
$\sim 8$ years [see Fig.~\ref{fig:bounds}].

\section{Amplification of quantum vacuum fluctuations}
\label{sec4}

The amplification of quantum vacuum fluctuations is a very general 
mechanism characterizing quantum field theory in curved space time, 
first discussed in cosmology by Grishchuk~\refcite{LG74} and 
Starobinsky~\refcite{AS79}. Let us first give a formal description of it.

We assume that the background field dynamics is described by the action:
\beq
S = \frac{1}{16\pi G_{\rm N}}\,\int d^{4} x \,\sqrt{|g|}\,{\mathcal R} 
+ S_m\,,
\eeq
where ${\mathcal R}$ is the Ricci scalar, 
$S_m$ refers to possible matter sources with $\sqrt{|g|}\,T_{\mu \nu} 
= 2 \delta S_m/\delta g^{\mu \nu}$, $T_{\mu \nu}$ being the energy-momentum 
tensor. We restrict ourselves to an isotropic and spatially homogeneous  
Friedmann-Lema\^{\i}tre-Robertson-Walker (FLRW) 
unperturbed metric, with scale factor $a$ and 
$ds^2 =g_{\mu \nu}\,dx^\mu\, dx^\nu=
-dt^2+a^2(t)\,d\vec{x}^2$. 
The free linearized wave equation for the TT  
metric perturbations $\delta g_{\mu \nu} = h_{\mu \nu}$ 
($h_{\mu 0}=0, \nabla_\mu h^{\mu}_{\,\,\nu}=0, h^\mu_{\,\,\mu}=0$) 
can be obtained by perturbing Einstein equations and keeping all 
the sources fixed, $\delta T_{\mu}^{\,\,\nu}=0$. 
A straightforward calculation gives: 
\beq
\Box h_{i}^{\,\,j}(t,\vec{x}) = 
\frac{1}{\sqrt{|g|}}\,\partial_\mu(\sqrt{|g|}\,g^{\mu \nu}\,\partial_\nu)\,
h_{i}^{\,\,j}(t,\vec{x}) =  0\,.
\eeq
Introducing the conformal time $\eta$, with $d\eta = dt/a(t)$ 
and writing 
\beq
h_{i}^{\,\,j}(\eta,\vec{x}) = \sqrt{8\pi\,G_{\rm N}}\,\sum_{P=+,\times}\,\sum_{\vec{k}} 
h^{\rm P}_{\vec{k}}(\eta)\,e^{i\vec{k} \cdot \vec{x}}\,\epsilon_{i}^{\rm P\,\,j}(\hat{\Omega})\,,
\eeq
each polarization mode $h^{\rm P}_{\vec{k}}(\eta)$ satisfies the equation
\beq
h^{\prime \prime}_{\vec{k}}(\eta) + 
2 \frac{a^\prime}{a}\,h^\prime_{\vec{k}}(\eta)+k^2\,h_{\vec{k}}(\eta)=0\,,
\label{eq:hpert}
\eeq
where we indicate with a prime the derivative with respect to 
conformal time $\eta$.

For simplicity let us consider two cosmological phases I and II, e.g.,
inflation and RD or MD phase.  We denote with $h_{\vec{k}}(\eta)$ and 
$H_{\vec{k}}(\eta)$ the solution of 
Eq.~(\ref{eq:hpert}) when the scale factor of phase I and II
is used, respectively. The mode expansion of $h_i^{\,\,j}$ in phase I reads:
\beq 
h_i^{\,\,j} = \sqrt{8 \pi\,G_N}\,\sum_{\rm P} \int \frac{d^3
k}{\sqrt{2 k}\,(2 \pi)^3}\, \left [ b_{\rm P}^{\rm
I}(\vec{k})\,h_{\vec{k}}(\eta)\,\epsilon_i^{\rm P\,\,j}(\hat{\Omega}) 
\,e^{i \vec{k}\cdot \vec{x}} + {\rm
H.C.} \right ]\,,
\label{hI}
\eeq
where H.C. stands for hermitian conjugate, $b_{\rm P}^{\rm I}(\vec{k})$ is
the annihilation operator with respect to the vacuum in phase I, 
i.e. $b_{\rm P}(\vec{k})|0\rangle_{\rm I}=0$ with ${\rm P} = \times,+$ [classically, it 
can be traded with a random variable.] We refer to $\vec{x}$ and 
$\vec{k}$ as the comoving coordinate and momentum, related   
to the physical quantities by $\vec{x}_{\rm phys}= 
a(t)\,\vec{x}$ and $\vec{k}_{\rm phys}= \vec{k}/a(t)$.
The mode expansion of $H_i^{\,\,j}$ in the phase II can be obtained from
Eq.~(\ref{hI}) with the substitution $h_{\vec{k}} \rightarrow
H_{\vec{k}}$, $b_{\rm P}^{\rm I}(\vec{k}) \rightarrow b_{\rm P}^{\rm
II}(\vec{k})$.  The annihilation operator of phase II satisfies 
the equations $b_{\rm P}^{\rm II}(\vec{k}) |0\rangle_{\rm II} =0$ 
with ${\rm P} = \times,+$.  
The quantities $\{h_{\vec{k}}, h_{\vec{k}}^*\}$ and $\{H_{\vec{k}}, 
H_{\vec{k}}^*\}$ are complete bases and can be related to each other by the so-called Bogoliubov
transformation~\refcite{BD}
\bea
\label{eq:Hbog}
&& H_{\vec{k}} = \sum_{\vec{k}^\prime} ( \alpha_{\vec{k} \vec{k}^\prime}\,
h_{\vec{k}^\prime} + \beta_{\vec{k} \vec{k}^\prime}\,
h^*_{\vec{k}^\prime})\,, \\
\label{eq:hbog}
&& h_{\vec{k}} = \sum_{\vec{k}^\prime} ( \alpha^*_{\vec{k} \vec{k}^\prime}\,
H_{\vec{k}^\prime} - \beta_{\vec{k} \vec{k}^\prime}\,
H^*_{\vec{k}^\prime})\,. 
\eea
Note that the above equations express the positive-frequency
components of the operator in phase I in terms of {\it both} the
positive and negative components of the operators in phase II, and
viceversa. The coefficients $\alpha_{\vec{k} \vec{k}^\prime}$ and
$\beta_{\vec{k} \vec{k}^\prime}$ are called Bogoliubov coefficients
and satisfy the equations~\refcite{BD}
\bea 
&& \sum_{\vec{k}} \left [ \alpha_{\vec{k}_1
\vec{k}}\,\alpha^*_{\vec{k}_2 \vec{k}} - \beta_{\vec{k}_1
\vec{k}}\,\beta^*_{\vec{k}_2 \vec{k}} \right ]=\delta_{\vec{k}_1
\vec{k}_2}\,, \\ && \sum_{\vec{k}} \left [ \alpha_{\vec{k}_1
\vec{k}}\,\beta_{\vec{k}_2 \vec{k}} - \beta_{\vec{k}_1
\vec{k}}\,\alpha_{\vec{k}_2 \vec{k}} \right ]=0\,.  
\eea
Inserting $H_{\vec{k}}$, given by Eq.~(\ref{eq:Hbog}), 
in the expression of $H_i^{\,\,j}$ 
and equating to $h_i^{\,\, j}$, we find the well-known 
relation between annihilation and 
creation operators in the two phases:
\bea 
&& b^{\rm I}(\vec{k}) = \sum_{\vec{k}^\prime} \left [
\alpha_{\vec{k}^\prime \vec{k}}\, b^{\rm II}(\vec{k}^\prime) +
\beta^*_{\vec{k}^\prime \vec{k}}\, b^{\rm II\,\dagger}(\vec{k}^\prime) \right
]\,,\\ && b^{\rm II}(\vec{k}) = \sum_{\vec{k}^\prime} \left [
\alpha^*_{\vec{k} \vec{k}^\prime}\, b^{\rm I}(\vec{k}^\prime) -
\beta^*_{\vec{k} \vec{k}^\prime}\, b^{\rm I\,\dagger}(\vec{k}^\prime) \right
]\,.  
\eea
Thus, if $\beta_{\vec{k} \vec{k}^\prime} \neq 0$, even starting from the vacuum 
${}_{\rm I}\langle 0|b^{{\rm I}\, \dagger}(\vec{k})\,b^{{\rm I}}(\vec{k}) 
|0\rangle_{\rm I} =0$, we end up with a final number of particles given by  
\beq
{}_{\rm I}\langle 0|b^{{\rm II} \dagger}(\vec{k})\,b^{{\rm II}}(\vec{k}) 
|0\rangle_{\rm I} = \sum_{\vec{k}^\prime} |\beta_{\vec{k}\,\vec{k}^\prime}|^2 
\neq 0\,.
\eeq
This phenomenon is known as {\it amplification of quantum-vacuum
fluctuations} in curved space time. It is due to the inevitable 
presence of positive and negative components in the 
basis I when expressed in terms of the basis II [see 
Eqs.~(\ref{eq:Hbog})]. The existence of the two complete bases, which are 
associated to two independent Fock spaces, is a consequence of being the 
space-time curved. 

The physical idea underlying the quantum-vacuum production of particles 
can be also explained as follows~\refcite{M00}. [Henceforth, we assume  
an isotropic and spatially homogeneous metric, so 
we can pose $\alpha_{\vec{k} \vec{k}^\prime}=\alpha_k\,
\delta_{\vec{k} \vec{k}^\prime}$ and $\beta_{\vec{k} \vec{k}^\prime}=\beta_k\,
\delta_{\vec{k} \vec{k}^\prime}$.] 
Suppose that over a time-scale $\Delta T \sim H^{-1}$ the cosmological 
phase changes, e.g., from phase I to phase II. If $t_*$ is the time at
which the transition occurs and $f_*$ is the physical frequency of the
mode at that time, there exist two possible regimes: abrupt transition
for which $2 \pi f_*\,H_*^{-1} \ll 1$ and adiabatic transition $2 \pi
f_*\,H_*^{-1} \gg 1$. Let us assume that the quantum state $|q \rangle$ 
before the transition has number of particles $N^{\rm I}_k 
\equiv \langle q| b_k^{\rm I\, \dagger }\,b_k^{{\rm I}} |q \rangle 
$. For modes for which the transition is sudden, i.e. $2 \pi
f_*\,H_*^{-1} \ll 1$ (for these modes 
$\lambda \hspace{-5pt}\raisebox{4pt}{\hbox{\tiny{/}}} 
\gg H_*^{-1}$, that is their wavelengths are larger than the Hubble radius or 
in jargon they are said to be {\it outside the horizon}, 
the physical state does not have time to change 
during the transition. However,
after the transition, the number of particles should be evaluated  
from annihilation and creation operators of phase II,
\beq 
N^{\rm II}_k\equiv \langle q| b_k^{{\rm II}\, \dagger}\,b_k^{{\rm
II}} |q \rangle = N^{\rm I}_k\,(1 + 2|\beta_k|^2) +
|\beta_k|^2\,.  
\label{prod}
\eeq
Therefore, if the number of particles when inflation starts is $N^{\rm I}_k$,  
this number is amplified by the factor $1 + 2|\beta_k|^2$~\refcite{A88,M00}. Moreover, 
due to the mixing between positive and negative 
frequencies even the vacuum state of phase I ($N^{\rm I}_k=0$) is a multiparticle 
state when referred to the vacuum state of phase II. 
On the other hand, for modes for which the transition is adiabatic, i.e. 
$2 \pi f_*\,H_*^{-1} \gg 1$ (for these modes $\lambda 
\hspace{-5pt}\raisebox{4pt}{\hbox{\tiny{/}}} \ll H_*^{-1}$, 
so their wavelengths are smaller than the Hubble radius  
or in jargon these modes are said to be {\it inside the horizon}), the quantum 
state has the time to follow the evolution of the scale factor. 
No particle production occurs in this case. In some sense the modes 
for which the transition is adiabatic, do not feel any effect of being the 
spacetime curved or changing in time. No mixing between positive and negative 
frequencies occurs, as it is in flat spacetime. 
[Practically, the relic GWs spectrum drops off exponentially 
for frequencies larger than $f_*^{\rm cutoff} \sim H_*/(2 \pi)$, i.e. 
for modes which never went outside the Hubble radius.]

We want now to give a more intuitive explanation of the 
phenomenon of amplification of quantum-vacuum fluctuations, which 
is very close to the original derivation due to Grishchuk~\refcite{LG74} 
and Starobinsky~\refcite{AS79}. It is based on a semiclassical approach.
By introducing the variable $\psi_k(\eta) = a\,h_k(\eta)$, $h_k$ being one of the two 
polarization modes, Eq.~(\ref{eq:hpert}) can be recast in the form
\beq
\label{eq:psi}
\psi_k^{\prime \prime} + \left [ k^2 - U(\eta) \right ]\,\psi_k =0\,,
\quad \quad U(\eta)= \frac{a^{\prime \prime}}{a}\,, 
\eeq
which is the equation of an harmonic oscillator in the time-dependent 
potential $U(\eta)$. [Note that to get unambiguous results, the change of variables 
$\psi(k) \rightarrow h(k)$ can be used only if the scale factor and 
its first derivative are continuous at the transition between 
phase I and II~\refcite{BMUV98}.] For simplicity, let us consider 
a de Sitter inflationary era, i.e. $a = -1/(\eta\,H_{\rm dS})$, 
$H_{\rm dS}$ being the constant Hubble parameter during de Sitter phase.

If $k^2 \ll |U(\eta)|$, since $|U(\eta)| \sim \eta^{-2}$ and $(a\,H_{\rm dS}) \sim 1/\eta$, 
then $k \eta \ll 1$, which means $k/a \ll H_{\rm dS}$ or $\lambda 
\hspace{-5pt}\raisebox{4pt}{\hbox{\tiny{/}}}\gg H_{\rm dS}^{-1}$, i.e. 
the mode is outside the Hubble radius (or in jargon under the potential barrier $|U(\eta)|$). 
In this case the solution of Eq.~(\ref{eq:psi}) is
\beq
\label{eq:sol1}
\psi_k \sim a\,\left [ A_k + B_k \,\int \frac{d \eta}{a^2(\eta)} \right ]\,\,\,
\rightarrow \,\,\,h_k \sim A_k + B_k\,\int \frac{d \eta}{a^2(\eta)}\,. 
\eeq
Since during de Sitter era the scale factor gets larger and larger,
the second term in the RHS of Eq.~(\ref{eq:sol1}) for $h_k$ 
becomes more and more negligible, thus the tensorial
perturbation $h_k$ remains (almost) constant while outside the Hubble radius. 
On the other hand, if $k^2 \gg |U(\eta)|$, we have $k\,\eta
\gg 1$ or $\lambda \hspace{-5pt}\raisebox{4pt}{\hbox{\tiny{/}}}
\ll H_{\rm dS}^{-1}$, which means the the 
mode is inside the Hubble radius (or in jargon over the potential barrier $|U(\eta)|$). 
The solution of Eq.~(\ref{eq:psi}) is a 
plane wave $\psi_k \sim e^{\pm i k\,\eta}$. Thus, in this case the
tensorial perturbation $h_k \sim e^{\pm i k\,\eta}/a$ decreases
in time while inside the Hubble radius. 
Therefore, in de Sitter case, the longer the tensorial-perturbation 
mode remains outside the Hubble radius, the more it gets amplified.
When the fluctuations re-enter the Hubble radius, the solution of 
Eq.~(\ref{eq:psi}) is 
\beq
\psi_k \sim \alpha_k e^{-i k \, \eta} + \beta_k e^{i k \, \eta}\,,
\eeq
and  contains both positive and negative modes. The coefficient 
$\beta_k$ is the Bogoliubov coefficient and it can be determined imposing 
the continuity of $\psi_k$ and $\psi_k^\prime$ all along the cosmological phases.

Before ending this section we want to express the intensity of the stochastic 
GW background (\ref{intGW}) in terms of the number of gravitons per cell 
of the phase space, which we indicate by $n_f$ with $f=|\vec{k}|/(2\pi)$. 
In the case of GWs produced out of vacuum, $n_f=N^{\rm II}_f$, thus it is the crucial quantity 
to evaluate. For an isotropic stochastic GW background $\rho_{\rm GW} = 
2 \int d^3k/(2\pi)^3\,(k\,n_k)$, thus 
\beq
\Omega_{\rm GW}(f) = \frac{1}{\rho_c}\,16\pi^2\,n_f\,f^4\,.
\label{spectbog}
\eeq
{}From the above expression it is straightforward to deduce that if the 
planned space- and ground- based GW experiments will detect a stochastic 
GW background whose spectrum satisfies the BBN phenomenological bound discussed 
in Section~\ref{sec3.4.1}, i.e. $\Omega_{\rm GW} < 10^{-6}$,
 the occupation number $n_f \gg 1$. This means the 
observed stochastic GW background is classical.

\subsection{Standard inflation}
\label{sec4.1}

In this section we shall first derive the GW spectrum in 
de Sitter inflation, which although is an ideal case (because it 
does not lead naturally to the transition from inflation to 
RD era), it has the advantage of providing analytical 
solutions for the tensorial perturbations.
Secondly, we discuss how the GW spectrum is modified in more 
realistic inflationary scenarios, like slow-roll inflation.

\subsubsection{de Sitter inflation}
\label{sec4.1.1}

The scale factor during de Sitter era is $a(\eta) = -1/(H_{\rm dS}\,\eta)$ 
with $-\infty < \eta < \eta_*$, while during the RD era we have 
$a(\eta) = (\eta - 2 \eta_*)/(H_{\rm dS}\,\eta_*^2)$ with $\eta_* < \eta < \eta_{\rm eq}$, 
where $\eta_{\rm eq}$ is the time at which there is equality  
between radiation and matter era. [Henceforth, we refer to quantities 
evaluated at present time with the subscript 0.]

Let us observe that during de Sitter phase any 
(classical) tensorial perturbation is {\it de-amplified} and their 
wavelength gets always stretched. Indeed, 
for a given comoving wavenumber $k$, it can be easily found~\refcite{BD01} 
that the de-amplification coefficient is $\mathcal{A}(f)=h(f,\eta_0)/h(f_{\rm in},\eta_{\rm in}) 
= e^{-\mathcal{N}}\,(H_{\rm dS}\,a_*/(H_0\,a_0))$, 
where $f=k/(2\pi\,a_0)$, $f_{\rm in}=k/(2\pi\,a_{\rm in})$ 
and $\mathcal{N}_{\rm in} \equiv \log (a_*/a_{\rm in})$ is the number of 
e-foldings. [To solve the {\it horizon} problem~\refcite{SC} the minimal 
amount of e-foldings is $\mathcal{N}_{\rm min} = \log (H_{\rm dS}\,a_*/(H_0\,a_0))$.] 
Thus, the longer the inflationary era, the smaller the coefficient 
$\mathcal{A}$. Moreover, the stretching of the wavelength is 
$\lambda/\lambda_{\rm in} = e^{\mathcal{N}}\,(a_0/a_*)$.

In de Sitter case Eq.~(\ref{eq:hpert}) can be solved exactly and the solutions 
are rather simple,~\footnote{These solutions can be obtained 
easily rewriting Eq.~(\protect\ref{eq:hpert}) in terms of 
the function $u_k$, defined by $h_k=(a\,u_k)^{\prime}/a^2$.} they read:
\bea
\label{f1}
&& h_k(\eta) = \frac{1}{a}\,\left ( 1- \frac{i}{k \eta} \right )\,e^{-i k\,\eta}\,,
\quad -\infty < \eta < \eta_* \\
&& h_k(\eta) = \frac{1}{a}\,\left [\alpha_k\,e^{-i k\,\eta} + \beta_k\,e^{i k\,\eta}
\right ]\,,
\quad \eta_* < \eta < \eta_{\rm eq}\,.
\label{f2}
\eea
[Here, we have assumed that the initial state of the Universe 
is the Bunch-Davis vacuum~\refcite{A88}, so in Eq.~(\ref{prod}) we could 
unambiguously set $N^{\rm I}_k=0$. See Ref.~\refcite{HK} where this assumption 
is relaxed.] 
The Bogoliubov coefficients $\alpha_k$ and $\beta_k$ in Eq.~(\ref{f2}) 
are obtained imposing the continuity of $h_k$ and $h^\prime_k$ across the transition.  A
straightforward calculation~\refcite{A88} gives $\beta_k =
1/(2k^2\,\eta_*^2)$, so the occupation number is $N_k = |\beta_k|^2$. 
Using the following relations: $k|\eta_*|=2\pi f\,a_0\,|\eta_*|=
2\pi\,f\,a_0/(a_*\,H_{\rm dS})$ where 
$a_* = 1/(H_{\rm dS}\,|\eta_*|)$, $a_0/a_* =
(t_0/t_{\rm eq})^{2/3}\,(t_{\rm eq}/t_*)^{1/2}$, we obtain 
$k|\eta_*| =f/f_*$ with 
\beq
f_* = \left (\frac{t_*}{t_{\rm eq}}\right )^{1/2}\,\frac{H_{\rm dS}}{2\pi}\frac{1}{1+z_{\rm eq}}=
10^9\,\left (\frac{H_{\rm dS}}{6\times 10^{-5}\,M_{\rm Pl}}\right )^{1/2} {\rm Hz}, 
\eeq
where 
we used $z_{\rm eq} \sim 10^4$, $t_{\rm eq} \sim 10^{10}$ s 
and $t_* = H_{\rm dS}/2$. The frequency $f_*$ is the cutoff frequency  
beyond which the GW spectrum falls down exponentially. 
By using Eq.~(\ref{spectbog}) with $n_k = 1/(4k^4\,\eta_*^4)$, it is 
straightforward to find that the GW spectrum for fluctuations that 
re-enter the Hubble radius during the RD era is then 
\beq 
\label{dS1}
h_0^2\,\Omega_{\rm GW}(f) \simeq 4 \times 10^{-14} \left (
\frac{H_{\rm dS}}{6 \times 10^{-5}\,M_{\rm Pl}} \right )^2\,.  \eeq
For fluctuations that re-enter the Hubble radius during the matter era, a similar 
calculation gives~\refcite{A88}:
\beq
\label{dS2}
h_0^2\,\Omega_{\rm GW}(f) \simeq 
4 \times 10^{-14}\,\left ( \frac{f_{\rm eq}}{f} \right )^2\, 
\left ( \frac{H_{\rm dS}}{6 \times 10^{-5}\,M_{\rm Pl}} \right )^2\,,
\eeq
where $f_{\rm eq} = H_{\rm eq}/(2 \pi)/(1+z_{\rm eq}) \sim 10^{-16}$ Hz. 
The COBE bound discussed in Section~\ref{sec3.4.2} imposes that the Hubble parameter 
during the de Sitter phase should be $< 6 \times 10^{-5}\,M_{\rm Pl}$ 
~\refcite{KW92}. 

\subsubsection{Slow-roll inflation}
\label{sec4.1.2}
\begin{figure}[t]
\begin{center}
\begin{tabular}{cc}
\hspace{-1.5cm}
\includegraphics[width=0.55\textwidth]{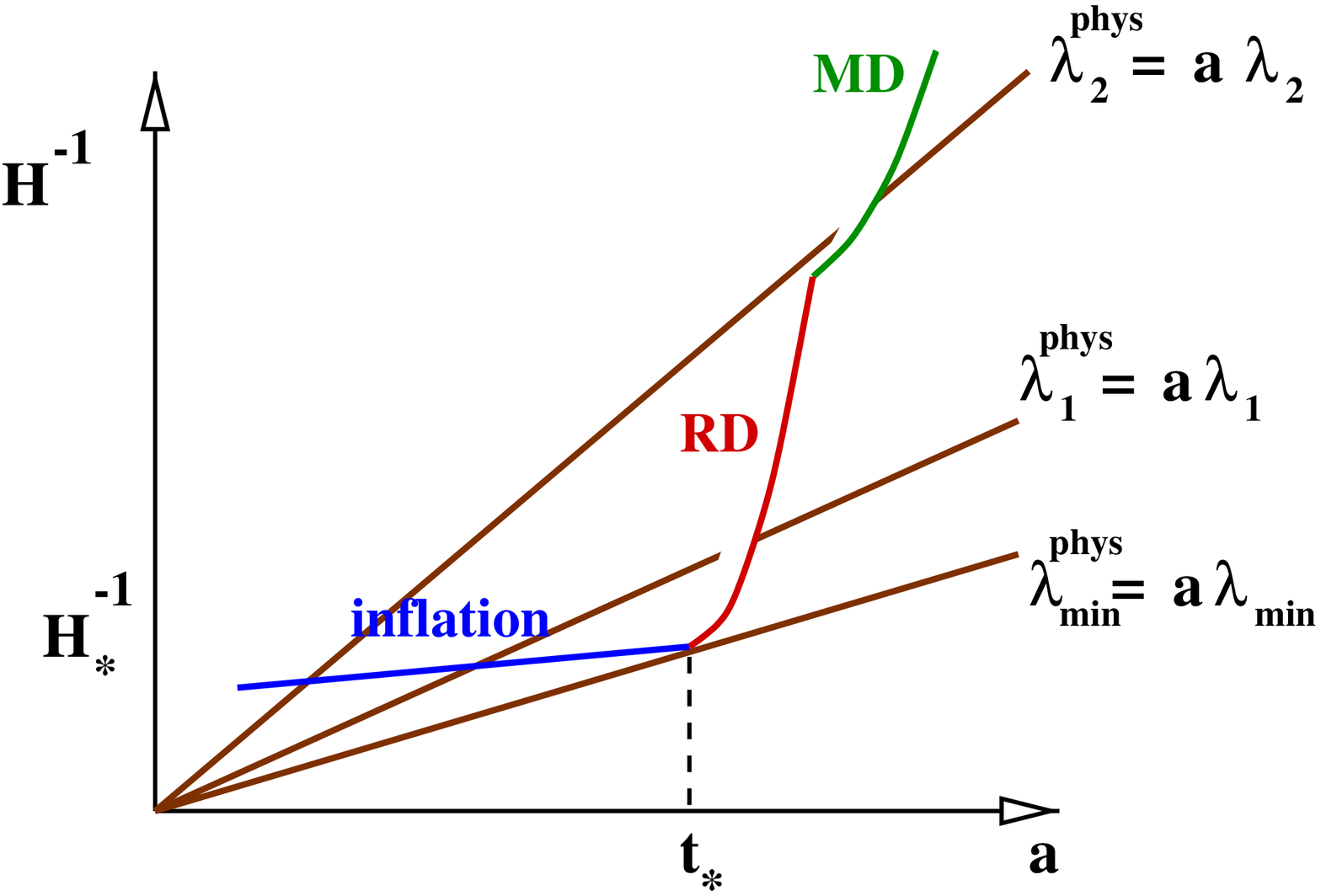}
&
\includegraphics[width=0.6\textwidth]{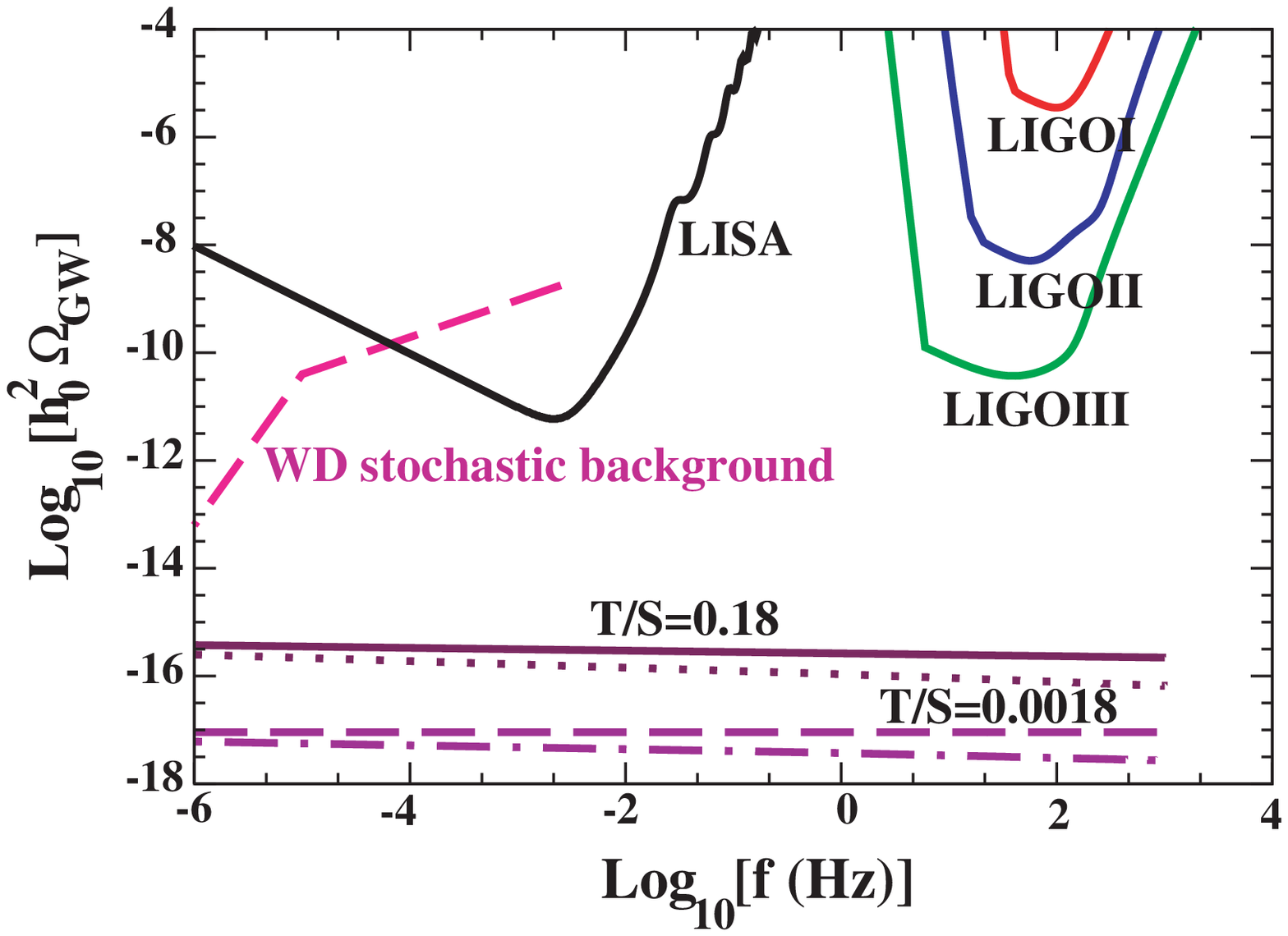}
\end{tabular}
\caption{\label{fig:standinfl} In the left panel we sketch the evolution 
of the expansion rate of the Universe $H$ and of some physical wavelengths 
during a period of standard inflation, followed by RD and MD eras. 
In the right panel we plot the stochastic GW background 
for slow-roll inflation obtained 
in Ref.~\protect\refcite{T96} for $T/S =0.18$ and $dn_T/d \log k = 0$ 
(continuous line), $T/S =0.18$ and $dn_T/d \log k = -10^{-3}$ 
(dot line), $T/S =0.0018$ and $dn_T/d \log k = 0$ (dash line) 
and $T/S =0.0018$ and $dn_T/d \log k = -10^{-3}$ (dot-dash line) 
The sensitivity of space-, (correlated) ground-based detectors and 
the stochastic background from WD binaries is also 
shown for comparison.}
\end{center}
\end{figure}
As a more realistic example we now discuss slow-roll inflation~\refcite{SC}. In this scenario 
a scalar field (the inflaton field) drives a period of accelerated expansion 
by rolling toward the minimum of its potential $V(\phi)$. [Inflation 
can occur if the slow-roll conditions are satisfied~\refcite{SC}: $\dot{\phi}^2 \ll 
V(\phi)$ and $|\ddot{\phi}| \ll |3H\,\dot{\phi}|, |V^\prime| \equiv |dV/d\phi|$.]
Differently from de Sitter inflation, in this case the Hubble parameter 
is not exactly constant during the inflation era, it slowly decreases in time, 
and as a consequence, the GW spectrum for fluctuations that go out of the 
Hubble radius during the inflation era and re-enter during RD era, is not flat 
but acquires a tilt: $h_0^2\,\Omega_{\rm GW}(f) \sim \overline{V}\,f^{n_{T}}$, 
\footnote{   
Whereas in de Sitter inflation the amplitude of the GW 
spectrum depends on the square of the (constant) Hubble parameter 
$H_{\rm dS}$ [see Eqs.~(\ref{dS1}), (\ref{dS2})], in potential-driven inflation, 
it depends on $\overline{V}$. Indeed, by applying the slow-roll conditions 
to the FLRW equations it can be easily seen that $H^2 \simeq 8\pi\,G_{\rm N}\,V/3$.}  
where we denote by $\overline{V}$ the value of the inflaton potential 
at the time the present Hubble scale crossed the Hubble radius during inflation.  
The spectral slope is~\refcite{T96} 
$n_{T} = -(\overline{V}^\prime/\overline{V})^2\,M_{\rm Pl}^2/(8\pi)$ and 
because of the slow-roll condition $|n_{T}| \ll 1$.
Quite nicely, the spectral slope can be expressed in terms of the scalar contribution $\mathcal{S}$ 
and tensorial contribution $\mathcal{T}\equiv 0.61 \overline{V}/M_{\rm Pl}$ 
to the quadrupole CMB anisotropy [see, e.g., Eqs.~(3), (4) in Ref.~\refcite{T96}].  
The result is $n_{T} = -\mathcal{T}/(7\mathcal{S})$. 

The explicit expression of the GW spectrum is given by Eqs.~(5), 
(6) in Ref.~\refcite{T96}. In Fig.~\ref{fig:standinfl} we plot the spectrum 
for two values of $\mathcal{T}/\mathcal{S}$, taking also into 
account the first-correction for the variation of the 
spectral slope with scale, $n_T(k)$, and compare 
it to the capabilities of LISA and (correlated \footnote{The 
curves have been produced by multypling LIGOI, LIGOII and LIGOIII sensitivities curves 
shown in Fig.~\protect\ref{ranges} by a constant factor obtained correlating 
Hanford (WA) and Louisiana (LO) LIGOs for four months 
at $90\%$ confident level assuming a flat GW spectrum.} )
ground-based detectors. Figure~\ref{fig:standinfl} shows that if the predictions of standard 
inflation are correct, unfortunately there is no hope of observing 
the stochastic GW background from slow-roll inflation 
with current and planned GW detectors. 

As seen from Fig.~\ref{fig:standinfl}, in the case of LISA, primordial 
GW backgrounds can be ``covered'' by the galactic WD binary 
background~\refcite{UV101}. Even if two LISAs will be used, 
by correlating their data stream, the minimum achievable 
sensitivity  is $\Omega_{\rm GW} 
\geq 10^{-13}$~\refcite{UV101}. Thus, to probe $\Omega_{\rm GW} 
\sim 10^{-16}\mbox{--}10^{-15}$ from inflation 
with space-detectors of LISA-kind, the detectors have to operate 
outside the regime of mHz which is dominated by close WD binaries. 
Various scientists have already started discussing a follow-on 
mission of LISA. Given the current astrophysical understanding, 
the most promising region should be $\sim 0.1 \mbox{--} 1$ Hz~\refcite{UV101,SKN,HB01,SP,KT}, 
which can be obtained by shortening LISA arms by a factor of $100$.
In this higher frequency band the astrophysically 
generated background from unresolved NS binaries is present; 
however, because the number of 
sources per frequency bin is expected to be not very high, it 
should be possible to remove it. To achieve sensitivities on the order 
of $\Omega_{\rm GW} \sim 10^{-16}\mbox{--}10^{-15}$, major, rather 
challenging improvements should be made, as increasing the laser power, 
the dimensions of the mirrors, etc..

The detection of relic GWs from inflation could be achieved 
also by measuring the polarization~\refcite{MZ}  in the CMBR 
with future CMB probes, as originally suggested in Ref.~\refcite{KKS}. 
Indeed, the polarization tensor on the celestial sphere 
can be decomposed into its ``gradient'' and ``curl'' components. 
The scalar fluctuations do not have {\it handness}, so they 
do not contribute to the curl, while 
tensorial fluctuations do contribute to the curl. 
If the polarization measurement will give a curl different from zero, 
GW will be detected and it will be possible to discriminate between the 
plethora of inflationary models currently available. Indeed,  
some of those models predict that the tensorial component 
$\mathcal{T}$ is negligible. 

\vspace{0.5cm}
In Sections~\ref{sec4.2}, \ref{sec4.3} we shall discuss other 
inflationary models. Before doing it, we want 
to point out that quite generally the tensorial perturbations 
will satisfy an equation similar to Eq.~(\ref{eq:psi}) but with the scale factor $a$ 
replaced by the so-called ``pump'' field $\tilde{a}$. The pump field depends  
on the dynamics of the scale factor and the other fields 
which are part of the background. Following Refs.~\refcite{GG93,G96,G97,BMUV98}, 
if we have two cosmological phases with $\tilde{a} = (\eta/\eta_*)^{\gamma_1}$ for 
$-\infty < \eta < \eta_1 <0$ and $\tilde{a} = |(\eta-2\eta_1)/\eta_1|^{\gamma_2}$ 
for $\eta_1 <\eta < \eta_2 $, then the solutions of Eq.~(\ref{eq:psi}) 
(having replaced $a$ with $\tilde{a}$) 
can be expressed in terms of Hankel functions of first and second species, as
\bea
\label{hank1}
\psi_k(\eta) &=& \sqrt{|\eta|}\,C\,H_{\nu_1}^{(1)}(k|\eta|)\,,\\
\label{hank2}
\psi_k(\eta) &=& \sqrt{|\eta-2\eta_1|}\,
[\alpha_k\,H_{\nu_1}^{(1)}(k|\eta-2\eta_1|) + 
\beta_k\,H_{\nu_1}^{(2)}(k|\eta-2\eta_1|)]\,,
\eea
where $\nu_1 = |\gamma_1 -1/2|$ and $\nu_2 = |\gamma_2 -1/2|$, and we have normalized 
Eq.~(\ref{hank1}) allowing only positive frequencies at $\eta \rightarrow - \infty$, 
so that $\psi_k \rightarrow Ce^{-ik\eta}/\sqrt{k}$. It can be derived 
that the Bogoliubov coefficient $\beta_k$ entering the GW spectrum is 
given by~\refcite{BMUV98} 
\bea
|\beta_f|^2 \sim f^{2\epsilon_T}\,\,\,  
&& \epsilon_T={1-|\gamma_1|-|\gamma_2|} \quad \mbox{for} \quad \gamma_1, \gamma_2 > 1/2 
\quad \mbox{or}\quad \gamma_1, \gamma_2 < -1/2\,, \nonumber \\
&& \epsilon_T={-|\gamma_1-\gamma_2|} \quad \quad \mbox{all\,\,other\,\,cases}\,.
\label{exp}
\eea
The spectral slope parameter $n_T$ is defined by 
\beq
n_T = \frac{d \log \Omega_{\rm GW}}{d \log f}=4+2\epsilon_T,
\eeq
and it is directly 
related to the kinematic behaviour of the background, that is to 
the evolution of the curvature scale and Hubble parameter. 
If we apply the above equations to de Sitter inflation, we find that 
$\gamma_1=-1$, for RD era $\gamma_2=1$ and for MD era $\gamma_2=2$. 
Thus, the spectral slope for fluctuations which re-enter the Hubble radius 
during RD era is $n_T=0$, and for those which re-enter the Hubble radius 
during MD era is $n_T=-2$, in agreement with Eqs.~(\ref{dS1}), (\ref{dS2}).
Moreover, in the case of slow-roll or power-law inflation, 
$\gamma_1 <-1$, and it can be easily found that for fluctuations 
that re-enter the Hubble radius during RD era $n_T =2+2\gamma_1<0$, as anticipated 
in Section~\ref{sec4.1.2}. 

Finally, we note that this way of estimating the spectral slope 
can be easily generalized~\refcite{G96} to the case of $n$ cosmological phases 
occurring at times $\eta_1, \eta_2, \dots$ if the pump 
fields follow a power-law evolution with coefficients 
$\gamma_1, \gamma_2, \dots$. In this case the relic GW spectrum is 
characterized by various frequency regions or {\it branches}, 
and in each of them the spectral slope depends {\it only} on the 
kinematics of the phases in which the mode associated to a given 
frequency went out of the Hubble radius and re-entered the Hubble radius.
On the other hand, the amplitude of the GW spectrum and special features 
in it, like possible oscillations, are determined by the 
dynamics and by physical assumptions, such as when and how the 
reheating process took place, whether entropy production is 
produced at some later stages, etc..

\subsection{Superstring-motivated cosmology}
\label{sec4.2}

Potential-driven inflation of the kind discussed in Section~\ref{sec4.1} 
cannot be easily implemented in string theory if the dilaton field 
or a modulus field is simply identified with the inflaton field~\refcite{CLO}.
This result forced people to conceive new ways of reconciling 
inflation and string theory. Henceforth, we shall briefly discuss 
some of those attempts which use supergravity description 
of superstring theory, the so-called pre--big-bang (PBB) 
scenario~\refcite{PBB} and bouncing-Universe 
scenarios~\refcite{KOST01,KOSST02}. Quite generally those 
scenarios predict a stochastic GW background anything but flat, 
which could be of interest for space- and ground-based detectors.

\subsubsection{Pre--big-bang scenario}
\label{sec4.2.1}

Due to the presence of other fields, at low energy 
superstring theory does not give Einstein general relativity 
-- for example heterotic string theory in four dimensions 
is described by the action
\beq
\label{eq:string}
\Gamma_{\rm eff} = \frac{1}{2\lambda_s^{2}}\,\int d^{4} x \,\sqrt{|g|}\,
e^{-{\varphi}}\,
\left [ {\mathcal R} + g^{\mu \nu}\,\partial_\mu {\varphi}\,
\partial_\nu {\varphi} - \frac{1}{12}(d B)^2 - V(\varphi)\right]\,,
\eeq
where $\varphi$ is the dilaton field, related to the string coupling by 
$g^2= e^{\varphi}$; $dB = \partial_\mu B_{\nu \rho} + 
\partial_\nu B_{\rho \mu} + \partial_\rho B_{\mu \nu}$, where $B_{\mu \nu}$ 
is the two-form gauge field or antisymmetric field; $V(\phi)$ is a 
non-perturbative potential; and where 
$\lambda_s$ is the string scale. In writing Eq.~(\ref{eq:string}) 
we have disregarded for simplicity the internal dimensions, whose dynamics 
can be described in terms of moduli fields~\refcite{PBB}.

Hencerforth, we limit the discussion to the homogeneous and isotropic case 
with $B=0$ and $V=0$ [$ds^2 = -dt^2 + a^2(t)\,d \vec{x}^2, 
\varphi= \varphi(t)$]. In this case the solution of the low-energy 
string-effective action (\ref{eq:string}) 
satisfies the scale-factor--duality  symmetry:
$a(t)\rightarrow 1/a(t)$, $\varphi(t) \rightarrow 
\varphi(t) - 6\log a(t)$,\footnote{Here for convenience the origin of time 
has been fixed at $t=0$.} 
with $a(t) \sim t^{1/\sqrt{3}}$ and $\varphi(t) \sim 
- \log t$. Noticing this property, Veneziano~\refcite{V91} conceived the idea of 
implementing the inflationary phase at times before the 
{\it would-be} big-bang singularity. Indeed, it is easily 
shown that for $t < 0$,  $\dot{a} >0, \ddot{a} >0$, thus   
the Universe undergoes a (super) inflationary phase.
Two different but physically equivalent 
descriptions of the PBB phase exist: either the string-frame  
picture described by Eq.~(\ref{eq:string}), where the Universe undergoes an 
accelerated expansion ($H > 0, \dot{H} >0, \dot{\varphi} > 0$), 
or the Einstein-frame picture, where the action (\ref{eq:string}) 
has the standard Hilbert-Einstein form and the evolution of the Universe 
is described by an accelerated contraction, or gravitational collapse 
($H < 0, \dot{H} < 0, \dot{\varphi} > 0$).

This new kind of inflation, which can be shown 
solves the homogeneity and flatness conundra, is driven by the 
kinetic energy of the dilaton field and forces 
both the string coupling ($\dot{g} >0$)
and the spacetime curvature to grow toward the future.
As a consequence, at least in the homogeneous case, the 
inflationary stage lasts for ever ($t \rightarrow - \infty$) and  
the initial state of the Universe is nearly flat, cold and decoupled: 
$g \ll 1$, ${\mathcal R} \lambda_s^2 \ll 1$.

The tensorial perturbations $h_k$ in the PBB model satisfy 
an equation similar to Eq.~(\ref{eq:hpert}) but with the 
scale factor $a$ replaced by the pump field 
$\widetilde{a} = e^{-\phi/2}\,a$, that is~\refcite{GG93}
\beq
h^{\prime \prime}_{\vec{k}}(\eta) + 
2 \frac{\tilde{a}^\prime}{\tilde{a}}\,h^\prime_{\vec{k}}(\eta)+k^2\,h_{\vec{k}}(\eta)=0\,.
\label{eq:hstring}
\eeq
Note that $\tilde{a}$ coincides with the scale factor in the Einstein frame. [In the case 
of $D=d+n+1$ dimensions with $n$ internal dimensions contracting   
and $d$ external dimensions expanding, isotropically, 
if tensorial perturbations depend only on the external 
dimensions, then Eq.~(\ref{eq:hstring}) is still valid but the pump field 
$\tilde{a} =e^{-\phi/2}\,a^{(d-1)/2}\,b^{n/2}$, $b$ being the scale factor 
of internal dimensions~\refcite{GG93}.] 
Let us consider the simplest scenario (henceforth minimal model) 
characterized by an isotropic  superinflationary phase 
with static internal dimensions, followed by RD era. In this case, 
the solution of the background field equations is~\refcite{PBB} 
$a(\eta) = (-\eta)^{(1-\sqrt{3})/2}$ and $\varphi(\eta) = 
\varphi_0-\sqrt{3}\,\log (-\eta)$, and during RD era, 
$a(\eta) = (\eta-2\eta_*)/(H_s\,\eta_*)$ 
and $\varphi(\eta) = \varphi_*$, where $H_s\simeq M_s=g_0\,
M_{\rm Pl}$, with $M_s$ the fundamental string mass 
and $g_0$ the string coupling at present time. 
Since during the superinflationary era the Hubble 
parameter is not constant ($\dot{H} \neq 0$), in fact it increases 
in time, we expect the GW spectrum for fluctuations re-entering 
the Hubble radius during RD era be non flat. Indeed, using Eq.~(\ref{exp}) 
with $\gamma_1 =1/2$ and $\gamma_2 = 1 $, it can be easily derived that 
the spectral slope is~\refcite{GG93,BGGV96,BMU97} $n_T=3$. If we assume that 
CMB photons we observe today carry the (red-shifted) 
energy of the primordial perturbations, i.e. 
$\Omega_\gamma(t_0)=(H_*/H_0)^2\,(a_*/a_0)^4$, 
and at the end of the superinflationary phase $H_*\simeq H_s$ and 
$g_*\simeq g_0$, then  the 
frequency at the end point of the spectrum is  
$f_* = k_*/(2\pi a_0) \simeq a_*\,H_s/(2 \pi a_0) \sim 10^{10}$ Hz, 
and~\refcite{GG93,BGGV96,BMU97}
\begin{figure}[t]
\begin{center}
\begin{tabular}{cc}
\hspace{-1.5cm}
\includegraphics[width=0.55\textwidth]{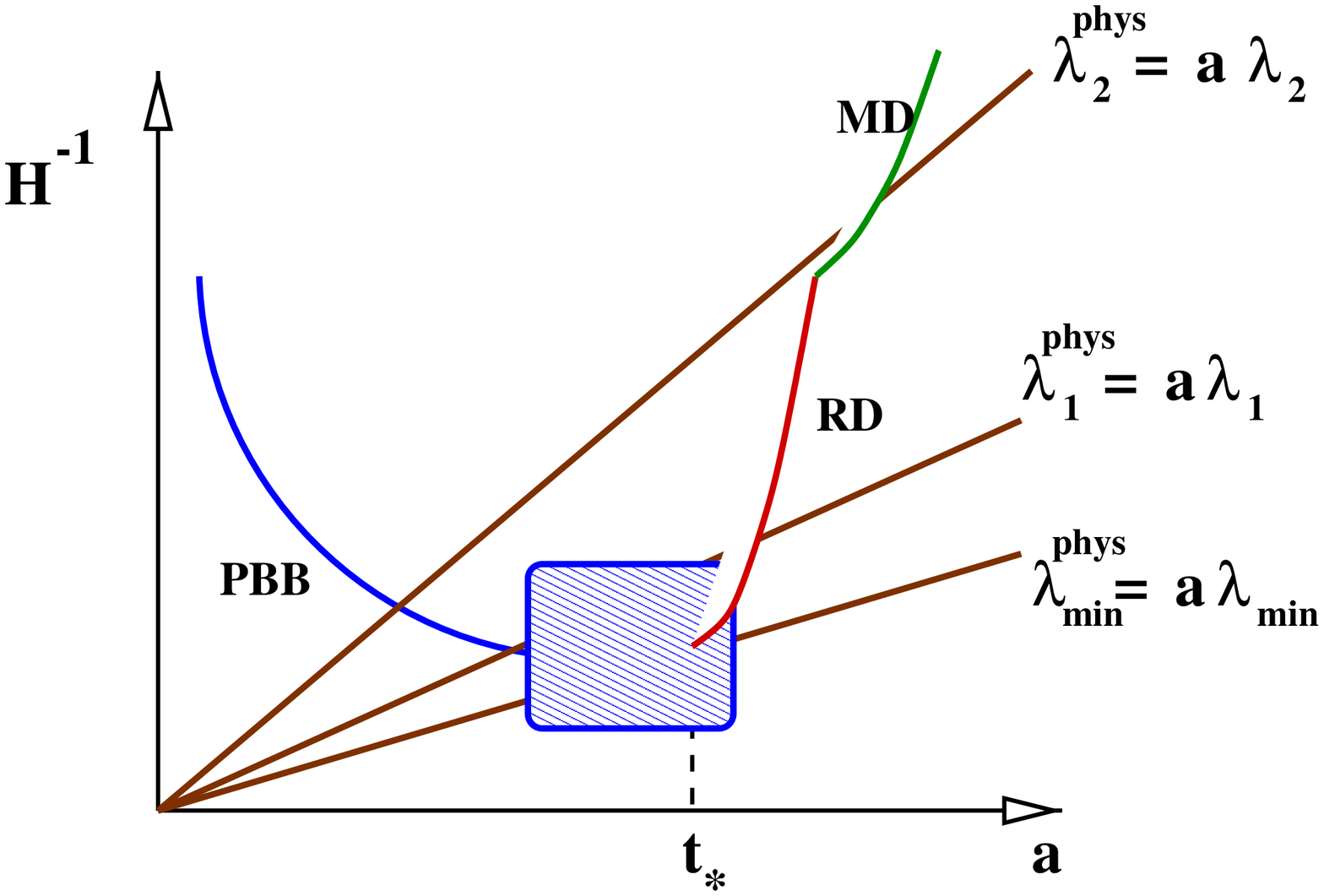}&
\includegraphics[width=0.6\textwidth]{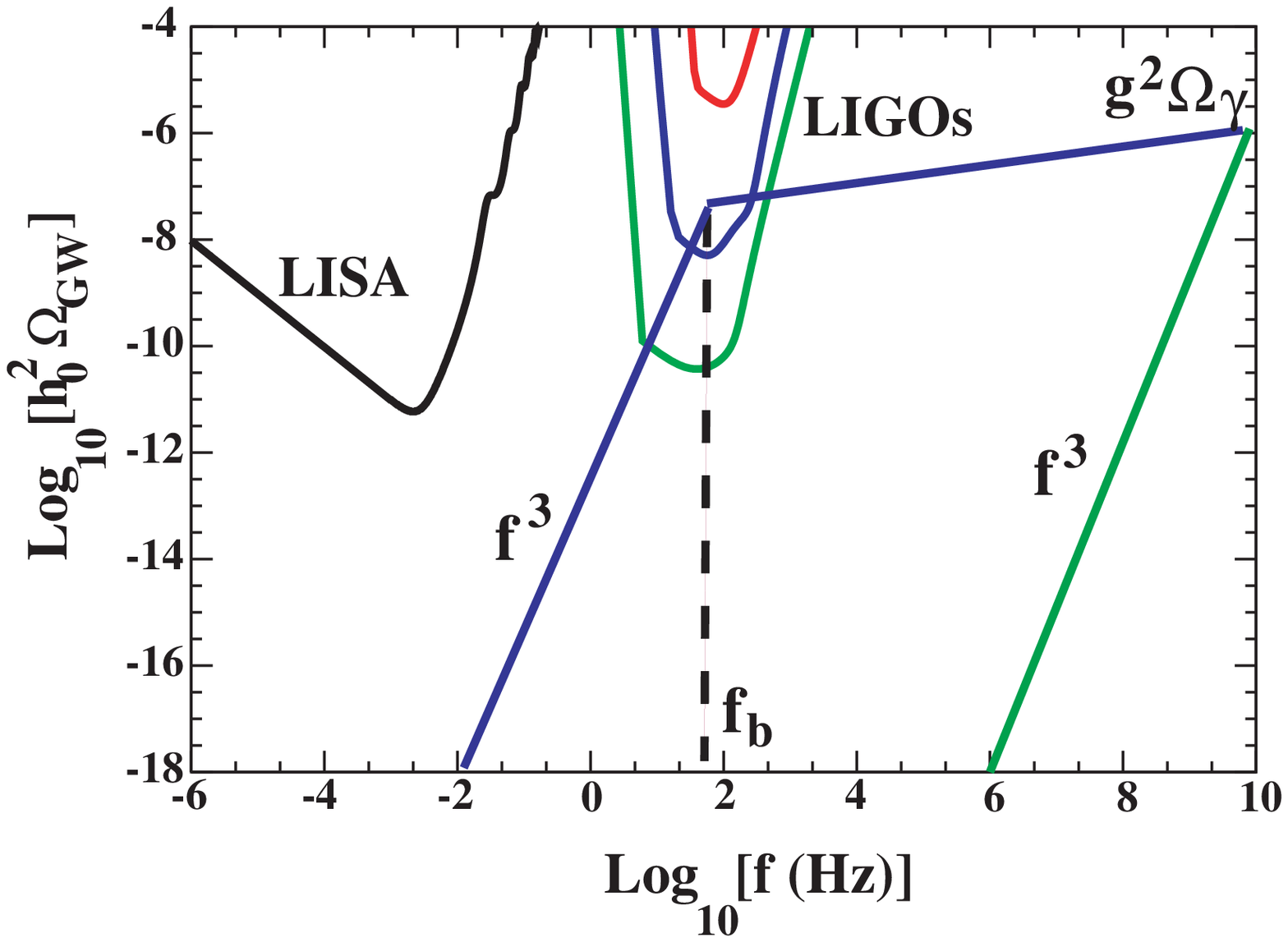}
\end{tabular}
\caption{\label{fig:pbb} In the left panel we sketch the evolution of 
$H$ and of some physical wavelengths during the superinflationary 
PBB phase, RD and MD eras. The shaded box refers to the spacetime
region around the would-be big-bang singularity for which 
we do not have a complete description, yet. In the right panel we show 
two examples of the stochastic GW spectrum.}
\end{center}
\end{figure}
\beq
\Omega_{\rm GW} = g_0^2\,\Omega_\gamma(t_0)\,\left (\frac{f}{f_*}\right )^3\,.
\eeq
Depending on the present value of the string coupling~\refcite{K85} $g_0 =0.03 
\mbox{--} 0.3$, the GW spectrum amplitude at the end point frequency 
is $\sim 10^{-7}\mbox{--}10^{-5}$ [$\Omega_\gamma(t_0) \sim 10^{-4}$], 
thus rather close to the BBN bound discussed in Section~\ref{sec3.4.1}.  
In the right panel of Fig.~\ref{fig:pbb} we show two examples 
of the stochastic GW background in the range of 
frequencies of space- and ground-based detectors. In the minimal 
model the spectrum has the peak in the GHz region and 
being elsewhere rather steep, it cannot be detected 
by current and planned GW detectors. Refs.~\refcite{BGGV96,BMU97,BMUV98} 
showed that if the PBB superinflationary phase is followed by 
other cosmological eras where (partial) higher-order derivatives and/or 
quantum loops corrections in the effective action (\ref{eq:string}) 
are included, the GW spectrum can have branches where it is 
less steeper, even flat, and could have amplitudes detectable 
by space- and ground-based detectors. However, robust 
predictions can be made only when a complete 
description of the transition from pre-- to the 
post--big-bang era will be available, including   
the understanding of the big-bang singularity. Indeed, the spectral 
slope for $f > f_{\rm b}$ in Fig.~\ref{fig:pbb}, depends 
crucially on the cosmological backgounds around the 
would-be big-bang singularity.

\vspace{-0.5cm}

\subsubsection{Bouncing-Universe scenarios}
\label{sec4.2.2}

New cosmologies motivated by string theory and influenced also 
by brane-world ideas have been proposed~\refcite{KOST01,KOSST02}. 
As in the PBB model, in those scenarios it is assumed that the 
Universe had a long existence prior to the big bang singularity.
Here, as an example, we shall focus on the bouncing-Universe model discussed 
in Ref.~\refcite{KOSST02}. This model uses two four-dimensional 
boundary branes, one of which is the visible brane where our Universe lives, 
the other is the hidden brane. The two branes are originally separated by a finite 
distance which can be regarded, according to the M-theory conjecture~\refcite{HW}, 
as the dilaton field in the strong-coupling regime. This scenario 
is built on the following four-dimensional effective action on the visible brane 
(in Einstein frame):
\beq
\label{eq:boun}
\Gamma_{\rm eff} = \frac{1}{16\pi G_{\rm N}}\,\int d^{4} x \,\sqrt{|g|}\,
\left [ {\mathcal R} - \frac{1}{2}g^{\mu \nu}\,\partial_\mu {\phi}\,
\partial_\nu {\phi} - V(\phi)\right]\,.
\eeq
The hidden boundary brane (or bulk brane) is assumed to be 
initially in a state with minimal energy, 
and it is flat, parallel to the visible brane and 
at rest. Non-perturbative effects, for example due to a  
potential of the kind $V(\phi) = -V_0\,e^{-{\rm c}\,\phi}$ 
with ${\rm c}>0$, can generate interactions 
between the visible and bulk brane, causing the 
bulk brane to move toward the visible one and collide
with it. In this way the kinetic energy of the bulk 
brane can be converted into radiation, and the hot big-bang 
era starts. 

Differently from the PBB scenario, in the model 
of Ref.~\refcite{KOSST02}, the Universe evolves from strong 
($g^2 = e^\phi \gg1$) to weak coupling ($g^2 = e^\phi \ll 1$), and both in the 
string- and Einstein-frame it undergoes an accelerated contraction. 
Moreover, it is assumed that going toward the singularity ($\eta \rightarrow 0^-$) 
the potential $V$ changes its shape and goes to zero.
Thus the cosmological background in the Einstein frame can 
be roughly described by a phase with $a_{\rm E}(\eta) 
\sim (-\eta)^\epsilon$ with $0<\epsilon \ll 1$ (very slow 
contraction), followed by a 
phase with (almost) zero potential, i.e. a superinflationary 
phase of the PBB kind, with $a_{\rm E}(\eta) \sim (-\eta)^{1/2}$ and then
a RD era. The pump field $\tilde{a}$ entering the equation of 
tensorial perturbations coincides with the scale factor 
in the Einstein frame ($\tilde{a} = a_E$), so for fluctuations which go out 
of the Hubble radius during the phase dominated by the potential and re-enter 
the Hubble radius during RD era, 
the discussion around Eqs.~(\ref{hank2})--(\ref{exp}) gives: 
$\gamma_1=\epsilon>0$ and $\gamma_2=1$, thus $n_T\sim 2+2\epsilon$, 
whereas for fluctuations 
that go out of the Hubble radius during the phase with (almost) zero 
potential and re-enter during RD era, we recover the PBB result $n_T=3$. 

\vspace{0.5cm}

Hence, both the PBB and bouncing-Universe scenarios predict very 
steep GW spectra at very low frequency, so no contribution of tensorial 
fluctuations to the CMBR inhomogeneities. 
Thus, if a tensorial component will be found in CMBR, 
then the current version of those scenarios should be rejected. 

Due to the collapse  phase, in both scenarios discussed in this section 
the initial (classical) tensor fluctuations are not de-amplified~\refcite{BD01}, 
as occurs in potential-driven inflation (see discussion at the beginning 
of Section~\ref{sec4.1.1}). This result, though paradoxical, does not imply 
that the initial value of tensor inhomogeneities must be 
fine tuned to unnaturally small value. Indeed, it can be derived~\refcite{BD01} 
that the energy density of tensor waves is indeed de-amplified. However, to 
solve the homogeneity problem in those superstring-inspired models, we have to apply 
constraints more severe~\refcite{BD01} than in potential-driven inflation models.

It is worth to mention that some aspects of both the PBB and bouncing-Universe 
scenarios, have been questioned. Doubts on the naturalness of the initial 
conditions and on the actual solution of the homogeneity and flatness conundra 
were raised~\refcite{IC}. The formation of large-scale structures 
and CMBR inhomogeneities cannot be explained in those models as naturally as 
in the standard inflationary models (see Section~\ref{sec3.1}) 
using adiabatic perturbations. 
[In the case of the PBB scenario, a way out has been 
recently suggested by converting isocurvature into adiabatic 
fluctuations~\refcite{BGGV} during the post--big-bang phase.] 
More importantly, a debate~\refcite{scalpert} on whether the scalar 
perturbations can be unambiguously calculated  
in those superstring-motivated models is still underway. 
Those issues can be clarified only when a complete and unique description 
of the transition from pre to post era will be available.

\subsection{Non-standard equations of state in post--big-bang eras}
\label{sec4.3}

In this section we want to give another example of 
non-flat relic GW spectrum due to the presence, soon 
after inflation, of a phase which differs 
from the RD era. The possibility of having an all variety 
of spectra by introducing non-standard, i.e. different from RD or MD, 
equations of state in post--big-bang 
eras was originally pointed out by Grishchuk~\refcite{LG74}.

\begin{figure}[t]
\begin{center}
\begin{tabular}{cc}
\includegraphics[width=0.6\textwidth]{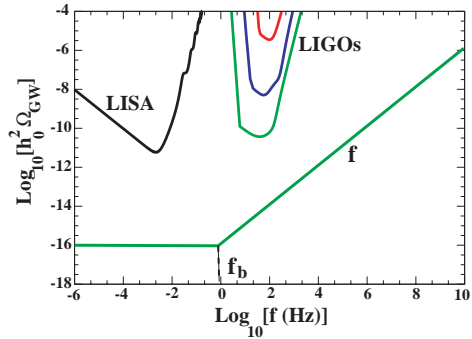}
\end{tabular}
\caption{\label{fig:quint} We show an example 
of the stochastic GW spectrum originated in quintessential 
inflation and contrast it with the sensitivity of space- and 
(correlated) ground-based GW detectors. For a detail discussion of the 
GW spectra as function of the free parameters see Ref.~\protect\refcite{G99}.}
\end{center}
\end{figure}

An interesting example is the case in which the inflationary 
phase is followed by an expanding phase driven by an 
effective source whose equation of state is stiffer than 
radiation. Peebles and Vilenkin~\refcite{PV99} discussed a model 
where the occurrence of a stiff post-inflationary phase 
is dynamically realized through a potential of the 
kind $V(\phi) = \lambda\,(\phi^4 + M^4)$ for $\phi < 0$ 
and $V(\phi) = \lambda\,M^8/(\phi^4 + M^4)$ for $\phi \geq 0$, 
with $\lambda \sim 10^{-14}$ and $M\sim 10^6$ GeV. 
This model has been denoted {\it quintessential inflation} and 
was originally introduced to explain the dark-energy conundrum, 
that is the present accelerated expansion of the Universe. 
The inflaton field 
starts its evolution at $\phi \ll -M_{\rm Pl}$ and rolls 
toward zero. Inflation ends at $\phi \sim -M_{\rm Pl}$ 
when significant part of the potential energy has turned 
into the kinetic energy of $\phi$. The cosmological evolution 
is then driven by the kinetic energy of $\phi$ (the 
equation of state is $p=\rho$), thus the scale 
factor $a(\eta) \sim \sqrt{\eta}$. Following the discussion 
around Eqs.~(\ref{hank2})--(\ref{exp}), we have $\gamma_2 = 1/2$, and  
assuming that during quintessential inflation~\refcite{G99}  
$a(\eta) \sim -1/\eta$, we have $\gamma_1 = -1$. So, the spectral slope 
for fluctuations that exit the Hubble radius during quintessential inflation and 
re-enter the Hubble radius during the 
stiff era is $n_T=1$~\refcite{PV99,G99}.  The GW spectrum increases linearly 
as a function of the frequency. An example of such spectrum is 
shown in Fig.~\ref{fig:quint}, 
where the flat branch refers to gravitons that go out of the Hubble radius 
during inflation and re-enter the Hubble radius during RD era. 
The peak of the spectrum is firmly localized at $100$ GHz, 
with amplitude $\leq\,10^{-6}$; the frequency $f_b$ at which the flat branch 
starts depend on the free parameters in the model~\refcite{G99}.   
A very detailed analysis which includes 
the discussion of the free parameters entering the spectrum and 
the comparison with current and planned GW detectors can be found 
in Ref.~\refcite{G99}. Unfortunately, in Peebles and Vilenkin 
quintessential model~\refcite{PV99} the spectrum is few order 
of magnitudes below the sensitivity achievable with 
advanced ground-based detectors. As shown in Ref.~\refcite{RU00},
by relaxing the assumption the scalar field driving the kinetic-energy 
phase coincide with the inflaton field, it is possible to shift the 
spectrum ending frequency at lower frequencies and increase 
by few orders of magnitudes the amplitude of the GW spectrum 
in the frequency range of ground-based detectors. 

\section{Gravitational waves from cosmic strings}
\label{sec5}

Topological defects could have been generated during 
symmetry-breaking phase transitions in 
the early Universe. Since the beginning of the 80s they received significant  
attention~\refcite{V85} as possible candidates for seeding structure formation. 
Recently, more accurate observations of CMBR inhomogeneities on 
smaller angular scales and compatibility with density-fluctuation spectrum 
on scales of 100 $h_0^{-1}$ Mpc, 
restrict the contribution of topological 
defects to at most $\sim 10\%$~\refcite{BPRS01}. 
In the following, we shall restrict our discussion to cosmic strings 
formed when a $U(1)$ local gauge symmetry is spontaneously broken [see Ref.~\refcite{ST02} 
for the possibility of producing cosmic strings from brane collision 
at the end of brane inflation scenario~\refcite{DT99}].

Cosmic strings are characterized by a single dimensional scale: 
the mass-per-unit-length $\mu$. Since they do not have any {\it ends}, 
they are in the form of loops, smaller or larger 
than the Hubble length. [The length of a string is defined as 
the energy of the loop divided by $\mu$.] 
Those loops have very large mass-per-unit-length, on the order 
of $10^{22} g/cm$ if the scale at which the symmetry is broken is 
the GUT scale. Their tension is equal to their mass-per-unit-length,   
so they oscillate relativistically emitting GWs and shrinking in size. 
It is generally assumed, though difficult to prove, 
that a network of cosmic strings did form at some time 
during the evolution of the Universe. In this network the 
only relevant scale is the Hubble length. Small loops 
(smaller than Hubble radius length) oscillate, 
emit GWs and disappear, but they are all the time replaced by small 
loops broken off very long loops (longer than Hubble radius length). The wavelength 
of the GW is determined by the length of the loop, and  
since in the network there are loops of all sizes, 
the GW spectrum is (almost) flat in a large frequency band, 
extending from $f\sim 10^{-8}$ Hz to $f\sim 10^{10}$ Hz.  
The GW spectrum is nearly  Gaussian and has been evaluated in detail 
in Refs.~\refcite{GWspec}. If $r$ is the characteristic radius of 
the loop, the quadrupole moment is $Q \sim \mu\,r^3$ and 
by assuming an oscillation period of $\tau \sim r$, 
the energy emitted is $d \mathcal{E}/dt \sim P \sim 
G_{\rm N}\,\tdot{Q}^2 \sim \gamma\,G_{\rm N}\,\mu^2$, 
where $\gamma$ is the dimensionless radiation efficiency $\sim 60$. 
Numerical simulations of cosmic-string network would predict 
$\Omega_{\rm GW}\sim P/\rho_c < 10^{-9}\mbox{--} 10^{-8}$ for 
cosmic strings with $G_{\rm N} \mu < 10^{-6}$. Thus, 
the Gaussian GW background from cosmic strings could 
be detectable by second and third generation of ground-based 
interferometers, and by LISA as well, being above the galactic stochastic 
GW background of WD binaries. 

More recently, Damour and Vilenkin~\refcite{DV01} found that strong beams of high-frequency 
GWs could be produced at cusps (where the string reaches 
a speed very close to light velocity) and at kinks along the string 
loop. [In time domain they are ``spikes'' or ``bursts''.]
As a consequence of these beams the GW background 
discussed above becomes strongly non-Gaussian. 
The most interesting feature of these GW bursts is 
that they could be detectable for a large range of values of 
$G_{\rm N} \mu$, larger than the usually considered search for the Gaussian spectrum.
For example, if the average number of cusps in a string oscillation 
is only $10\%$, the GW bursts could be detectable by 
ground- and space-based detectors up to $G_{\rm N} \mu \sim 10^{-13}$.  
If the number of cusps in a string oscillation is very small, kinks, which have 
smaller bursts but are ubiquitous, could generate a signal 
detectable by LISA for a large 
range of values of $G_{\rm N} \mu$ [see Figs.~1, 2 in Ref.~\refcite{DV01}].

The most stringent bound on the GW backround from cosmic strings comes 
from pulsar timing [see Section~\ref{sec3.5.3}]. The analysis of Ref.~\refcite{DV01} 
suggests that the GUT value $G_{\rm N} \mu_{\rm GUT} \sim 10^{-6}$ is still compatible 
with existing pulsar data, and if accuracy will be improved,  
pulsar timing could allow to detect the GW bursts 
up to $G_{\rm N}\mu \sim 10^{-11}$.

\section{Gravitational waves from other mechanisms occurring in early Universe}
\label{sec6}

During its history the Universe could have undergone phase transitions, 
corresponding to symmetry breaking of particle-physics fundamental 
interactions. 

In a first-order phase transition the Universe is initially trapped 
in a high-temperature metastable phase (unbroken-symmetry phase), 
where the {\it false} vacuum 
is separated from the {\it true} vacuum by a barrier in
the potential $V(\phi,T)$ of some scalar field $\phi$ driving the
transition ($\phi$ is also called the order parameter) [see
Fig~\ref{fig:FOPT}]. The transition from the metastable to the ground state
takes place via quantum tunneling across the barrier with  
random nucleation of bubbles: true vacuum is inside the bubbles, 
false vacuum is outside them.  
If the temperature when bubbles form is too high, they are born small, the 
volume energy cannot overcome the shrinking effect of the surface
tension and they disappear quickly; but when the temperature drops below a
critical temperature, bubbles are born larger, and since the
latent heat released during the transition is converted to bubble-wall 
kinetic energy, they expand, approach velocity close to the speed 
of light, collide and leave the Universe in 
the broken-symmetry phase.  When the collision occurs, the spherical
symmetry is broken and GWs, as well as other particles~\refcite{WH}, are
radiated away~\refcite{BubbleColl}. Since the Universe is expanding, 
the temperature $T_*$ at which the transition takes place can be obtained
comparing the probability of bubble nucleation per unit time and
volume with the expansion rate of the Universe at that temperature.
The transition occurs when the probability for a single bubble to be
nucleated within one horizon volume is on the order of one.  

Two parameters determine the stochastic GW spectrum~\refcite{KKT94}: the bubble
nucleation rate per unit volume $\beta$ ($\Gamma =
\Gamma_0\,e^{\beta\, t}$) and the ratio $\alpha$ between the false
vacuum energy density and the energy density of the radiation at the
transition temperature $T_*$.  The bubble collision produces a 
GW spectrum which is strongly peaked at the frequency characteristic of
the nucleation rate, i.e. $2 \pi f_{\rm peak}
\simeq \beta$.  A detailed analysis gives~\refcite{KKT94}:
\begin{figure}[t]
\begin{center}
\includegraphics[width=0.5\textwidth]{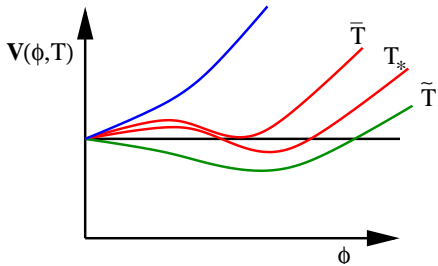}
\caption{We show the typical temperature-dependence 
of a potential $V(\phi,T)$ in a first-order phase transition 
$\phi$ being the order parameter. At the temperature 
$\overline{T}$ the true ($<\phi>\neq 0$) and false (conventionally $<\phi>= 0$) 
vacua are degenerate, and at $\tilde{T}$ the false vacuum becomes unstable. 
The transition occurs at some temperature $\tilde{T}<T_\star<\overline{T}$ \label{fig:FOPT}}
\end{center}
\end{figure}
\beq
f_{\rm peak} \simeq 5.2 \times 10^{-8}\,\left ( 
\frac{\beta}{H_*}\right )\,\left ( 
\frac{T_*}{1\, {\rm Gev}}\right )\,
\left (\frac{g_*}{100}\right )^{1/6}\,{\rm Hz}\,,
\label{peakf}
\eeq
where $H_*$ and $g_*$ are the Hubble parameter and the number of degrees 
of freedom at the time of transition. Typical values 
for an electro-weak phase transition (EWPT) are 
$\beta/H_* \simeq 10^2 \mbox{--} 10^3$, 
$T_* = {\mathcal O}(100)$ GeV, for which 
$f_{\rm peak} \simeq 10^{-4}\mbox{--} 5\times 10^{-3}$ Hz, 
and lies in the LISA frequency band. The GW 
spectrum was calculated in Ref.~\refcite{KKT94}; it reads
\beq
h_0^2\,\Omega_{\rm GW} \simeq 10^{-6}\,\kappa^2\,\frac{\alpha^2}{(1+\alpha)^2}\,
\frac{v_b^3}{0.24 + v_b^3}\,\left ( 
\frac{H_*}{\beta}\right )^2\,
\left (\frac{100}{g_*}\right )^{1/3}\,,
\label{peakGW}
\eeq
where $\kappa$ quantifies the fraction of latent heat that is transformed 
into bubble-wall kinetic energy and $v_b$ is the bubble expansion velocity. 
At frequencies lower than $f_{\rm peak}$ the 
GW spectrum $\propto f^{2.8}$~\refcite{KKT94}, 
while at higher frequencies it drops off 
as $\propto f^{-1.8}$~\refcite{KKT94}.

Non-perturbative calculations done using lattice field theory 
have shown that there is no first-order EWPT in the SM of particle physics 
for Higgs mass larger than W masses (current results 
predict an Higgs mass larger than W masses).  
In Minimal Supersymmetric Standard Models (MSSM), if the Higgs mass is in the range 
$110\mbox{--}115$ GeV, there is the possibility of having a first-order 
phase transition, if the right-handed stop mass has  
a mass in the range $105\mbox{--}165$ GeV [the stop particle is 
the scalar supersymmetric partner of the top quark]. However, 
in this case the GWs produced are too weak. 
For example, for a Higgs mass of 110 GeV and right-handed stop mass 
of 140 GeV, there exist regions in the parameter space 
where~\refcite{AMNR01} 
$\alpha \sim 10^{-2}$, $\kappa \sim 0.05$ and $H_*/\beta \simeq 10^{-4}$ and 
thus $h_0^2\,\Omega_{\rm GW} \sim 10^{-19}$ around $10$ mHz. 
The strength of the transition can be enhanced in next-to-minimal 
supersymmetric standard models (NMSSM). 
Apreda et al.~\refcite{AMNR01} investigated the NMSSM obtained adding a gauge singlet 
in the Higgs sector~\refcite{EGHRZ89}. This model is 
rather attractive also because it can explain the observed baryon asymmetry. 
Apreda et al. found that there exist regions in the 
parameter space for which the GW spectrum is  
$h_0^2\,\Omega_{\rm GW} \sim 10^{-15} \mbox{--}10^{-10}$ 
at $f_{\rm peak} \simeq 10$ mHz. Note that for frequencies 
$10^{-4}\mbox{--}3 \times 10^{-3}$ Hz the stochastic GW background 
from WD binaries could ``cover'' the GW spectrum from first-order 
phase transitions. If this GW background will be ever detected 
it will be a signature of supersymmetry and, combined with experimental 
bounds from future particle colliders, it could allow to 
discriminate between various supersymmetric models. 

A stochastic GW background could be also produced during a first-order
phase transition from turbulent (anisotropic) eddies generated  
in the background fluid during the fast expansion and collision 
of the true-vacuum bubbles~\refcite{KKT94,AMNR01,KMK02}. In the 
NMSSM~\refcite{EGHRZ89} there exist regions 
of the parameter space where~\refcite{AMNR01} $h_0^2\Omega_{\rm GW} \sim 10^{-10}$ with 
peak frequency in the mHz. Recently, the authors of Ref.~\refcite{DGN02} 
evaluated the stochastic GW background generated by 
cosmic turbulence before neutrino decoupling, i.e. much later than 
EWPT, and at the end of a first-order phase transition 
if magnetic fields also affect the turbulent energy spectrum. 
The observational perspectives of those scenarios are 
promising for LISA.  

Turner and Wilczek~\refcite{TW90} pointed out that if inflation 
ends with bubble collisions, as in extended inflation, 
the GW spectrum produced has a peak in the frequency range 
of ground-based detectors. Subsequent analyses have shown that
in two-field inflationary models where a field 
performs the first-order phase transition and a second field 
provides the inflationary slow rolling (so-called first-order 
or false vacuum inflation~\refcite{FOI}), if the phase transition 
occurs well before the end of inflation, a GW spectrum peaked 
around $10 \mbox{--} 10^3$ Hz, can be produced~\refcite{BAFO97}, with an amplitude 
large enough, depending on the number of e-foldings left after the 
phase transition, to be detectable by ground-based 
interferometers. A successful detection of such a spectrum 
will allow to distinguish between inflation and other cosmological 
phase transitions, like QCD or electroweak, which have a different peak frequency. 

Another mechanism that could have produced GWs in the early universe 
is parametric amplification after preheating~\refcite{KT97}. During 
this phase classical fluctuations produced by the oscillations of 
the inflaton field $\phi$ can interact back, via parametric resonance, 
on the oscillating background producing GWs. In the model where 
the inflaton potential contains also the 
interaction term $\sim \phi^2\,\chi^2$, $\chi$ being a scalar field,  
the authors of Ref.~\refcite{KT97} estimated 
$\Omega_{\rm GW} \sim 10^{-12}$ at $f_{\rm min} \sim 10^5$ Hz, while in 
pure chaotic inflation $\Omega_{\rm GW} \,\leq\, 10^{-11}$ at 
$f_{\rm min} \sim 10^4$ Hz. [See Fig. 3 in Ref.~\refcite{KT97} 
for the GW spectrum in the range $10^6\mbox{--}10^8$ Hz.]
Unfortunately, the predictions lie in frequency range where no GW detectors 
have been planned so far, although EM cavities to detect GWs 
were proposed~\refcite{EMRussians}.

\section{Gravitational waves in brane world scenarios}
\label{sec7}

During the last years there have been feverish activities around 
brane-world models~\refcite{BW,CC}. Those scenarios are based on the idea 
that large ($\geq 1/M_{\rm Pl}$) spatial, extra dimensions might exist. 
They assume that ordinary matter is confined onto a three-dimensional 
subspace, called the (visible) {\it brane}, which is embedded in 
a larger space, called the {\it bulk}. Only gravitational 
interactions are allowed to propagate in the bulk. 
If $n$ is the number of extra dimensions,  
then the four-dimensional Planck mass $M_{\rm Pl}$ is 
a derived quantity, while the fundamental scale is determined 
by the gravitational mass $M_{\rm fund}$ in $n+4$ dimensions. 
In the case of a flat bulk, $l$ being the typical length of 
the extra dimensions, $M_{\rm Pl}^2 = M_{\rm fund}^{n+2}\,l^n$ 
with $M_{\rm fund}$ ranging between TeV to $M_{\rm Pl}$.  
In the following, we shall mention results obtained in five-dimensional 
brane-world models ($n=1$). 

By normalizing properly Eq.~(\ref{f1}) and taking the limit 
$k \eta \ll 1$, it is easily derived 
that long-wavelength GWs generated in de Sitter inflation have a 
(almost) constant amplitude $k^{3/2}\,h_k = H_{\rm dS}/M_{\rm Pl}$, 
which depends on the two scales in the problem: the Hubble 
parameter during inflation and the low-energy Planck mass. 
In five-dimensional brane-world models 
there is an additional scale in the problem, the scale $l$ 
associated to the fifth dimension. Quite generically, we could 
expect~\refcite{FK02} that if inflation on the (visible) brane occurs 
at scales smaller than the curvature of the extra dimension, 
i.e. $H\,l \gg 1$, then the amplitude of GWs~\footnote{Here 
the GW refers to the homogeneous Kaluza-Klein 
massless mode. Massive Kaluza-Klein gravitons decay very fastly~\refcite{LMW00,GRS01,FK02}, 
so their production is very suppressed during brane inflation.} 
can be modified with respect to the value 
predicted in standard four-dimensional theory -- for example~\refcite{LMW00,GRS01,GKLR02,FK02}  
$k^{3/2}\,h_k = f(H\,l)\,H/M_{\rm Pl}$ with some function $f$.~\footnote{In the models 
analysed by Frolov and Kofman~\refcite{FK02}, $f(H\,l)/M_{\rm Pl} =1/M_{\rm Pl, infl}$ where 
$M_{\rm Pl, infl}$ is the Planck mass during inflation.} 
In the brane-world models analysed in Refs.~\refcite{LMW00,GRS01,FK02}, it 
was found that for $H\,l \gg 1 $ the tensor amplitude is enhanced with respect to 
the standard result, i.e. $f(l\,H) >1$. Moreover, 
the ratio between scalar and tensorial perturbations 
$\mathcal{T}/\mathcal{S}$ may also differ from the value
predicted by standard four-dimensional theory~\refcite{LMW00,FK02} and 
it could depend on specific features of the brane-inflation model.
However, as seen in Section~\ref{sec3}, the overall shape of the 
relic GW spectrum depends not only on the evolution of the 
tensor-mode amplitude during inflation but also on the post--big-bang 
phases, i.e. on the era at which the mode re-enters 
the Hubble radius. Only when those cosmological phases 
will be consistently described in the brane-world scenarios, 
we will have robust predictions for the GW spectrum. 

By implementing quintessential inflation on the visible brane, and evaluating the Bogoliubov 
coeffients all along the Universe evolution, the authors 
of Ref.~\refcite{SSS01} found results similar to the ones discussed in Section~\ref{sec4.3}: 
the GW spectrum has a branch where it 
increases linearly as function of the frequency. 

GWs can be also produced by excitations of the 
so-called radion field, which controls the size of the extra 
dimension, and by inhomogeneities in the displacement of 
the brane~\refcite{CH00}. Those waves should peak at a frequency fixed by the scale 
of the extra dimension. LISA could observe excitations emitted at scales 
from millimeters to microns, while ground-based interferometers 
could observe rather small extra dimensions, up to $\sim 10^{-12}$ mm.

In some brane world scenarios the causal propagation of 
gravitational and luminous signals can be different~\refcite{all,CEG01}. 
In the case of an asymmetric warped spacetime of the kind~\refcite{CEG01},
\beq
ds^2 = -n^2(l)\,dt^2 + a^2(l)\left ( \frac{d r^2}{1-k\,r^2}
+ r^2\,d \Omega_2^2 \right ) + b^2(l)\,dl^2\,,
\eeq
where $l$ is the coordinate along the extra dimension, 
and $k=\pm 1,0$ is the spatial curvature of the 
three-dimensional sections parallel to the brane,   
the local speed of light $c(l)=n(l)/a(l)$ depends on the 
position $l$ along the extra dimension. Since 
GWs can propagate in the bulk, a GW signal emitted at point 
$\mathcal{A}$ on the brane can take [if $c(l)$ is increasing away 
from the brane] a {\it short-cut} in the bulk, and 
to an observer located at point $\mathcal{B}$ on the brane 
it appears quicker than a photon traveling  
along the brane from $\mathcal{A}$ to $\mathcal{B}$. 
If true, this effect predicts a difference between 
gravitational and luminous speed. By detecting GWs and EM waves 
from $\gamma$-ray bursts and supernovae, ground-based detectors 
of second generation can put upper limits on $\delta c/c$ 
at most on the order of $10^{-17}$~\refcite{CT02} for $\gamma$-ray 
bursts, and $10^{-11}$ for a supernova in the Virgo cluster of galaxies.

\section{Extraction of cosmological parameters from detection of GWs}
\label{sec8}

In this section we want to discuss very briefly another possible 
way GW detection can be used to probe cosmology, although not 
directly the physical mechanisms occurring in the very early Universe.
 
It was realized long ago that binaries made of compact bodies, 
like BHs or NSs, which spiral in toward each other loosing 
energy because of the emission of GWs, are ``standard candles''~\refcite{S85}.
Indeed, once the binary masses and spins, position and 
orientation angles are specified, the GW signal 
passing-by the detector depends only on the luminosity distance 
$d_{\rm L}(z,\Omega_M,\Omega_\Lambda,\cdots)$ 
between binary and detector. By using three ground-based 
interferometers or LISA (and suitably high signal-to-noise ratio) it 
is possible to determine the location and orientation 
of the binary in the sky, extract the masses, the spins and the 
cosmological distance but not the source's cosmological red-shift. 
Indeed, the inspiral GW signal emitted from a binary with masses 
$(m_1,m_2)$ at red-shift $z$ cannot be distinguished, 
except for an overall factor in the amplitude, from the one 
emitted from a local binary with masses $[(1+z)\,m_1,(1+z)\,m_2]$.
To break the degeneracy and obtain the distance--red-shift 
curve and the cosmological parameters, one could associate 
the binary-coalescence event to an EM event which had very clear emission or  
absorption lines, from which one could read $z$. 
[Because the errors in position determinations can be 
rather large, the existence of EM counterpart can also improve 
the accuracy in measuring  the luminosity distance -- 
for example by mapping massive BH binaries with LISA 
$\delta d_{\rm L}/d_{\rm L}$ decreases from
$1\mbox{--} 10 \%$ to $ 0.15 \mbox{--} 1\%$~\refcite{HH02}.]
However, Holz and Hughes~\refcite{HH02} recently pointed out that, practically, 
because of gravitational-lensing effects 
(GWs are lensed as EM radiation are lensed) the precision 
with which the luminosity distance can be determined is 
degraded, though still slightly better than what Type-Ia supernovae 
probes can reach, but unless the event rates is high enough 
the extraction of cosmological parameters cannot be done 
with good accuracy.

\section{Summary}
\label{sec9}

As seen, the search for GWs from the very early Universe is rather 
challenging but the outcome is certainly worth the effort. 

Theoretically, so little we know about the very early evolution of 
our Universe, that the predictions for the stochastic GW background 
from standard inflation~\refcite{LG74,AS79} and/or superstring-motivated 
models~\refcite{PBB,KOST01}, including also brane-world scenarios
~\refcite{LMW00,GRS01,GKLR02,SSS01,CH00}, should be considered just as indications. 
What we learned from those models, is that it could well be the 
GW spectrum is not (almost) flat all the way from $f \sim 10^{-16}$ Hz to $f\sim 10^{10}$ Hz. 
There could be frequency regions in which it increases or 
decreases~\refcite{LG74} due to cosmological phases existed before the would-be 
big-bang singularity, 
and/or post--big-bang phases with equations of state different than radiation or matter ones, 
whose presence we cannot currently exclude. 
[Even an initial state different from vacuum could affect the relic GW spectrum~\refcite{HK}, 
as seen from Eq.~(\ref{prod}).] Future CMB polarization experiments~\refcite{KKS} could 
detect the tensor contribution at very large wavelength 
and discriminate within plethora of inflationary models.
The detection of GWs from bubble collision~\refcite{BubbleColl,AMNR01,FOI,BAFO97} and/or  
turbulent motion~\refcite{BubbleColl,AMNR01,KMK02,DGN02} in the primordial plasma, 
and especially the frequency at which the GW spectrum 
is peaked, will reveal if first-order phase transitions occured, and at which 
particle physics mechanism they are associated to: QCD, EW, last stages of 
inflation, etc.  
If a network of 
cosmic strings ever formed, it should have produced a Gaussian~\refcite{GWspec} or  
even strongly non-Gaussian~\refcite{DV01} GW spectrum,  which could be 
detectable by the second generation of ground-based interferometers and LISA. 
If not observed, we could put upper limits on few free parameters. 
An independent estimation of the cosmological parameters could be 
obtained by detecting GWs from coalescing binaries at fairly high red-shift
~\refcite{S85,HH02}.

Experimentally, since the GW signal from astrophysical sources 
is much more promising than the one from the early Universe, 
the ground- and space-based detectors were conceived and 
designed to detect the formers. Various scientists in the GW community have 
already started thinking at future detectors, 
like the follow-on LISA missions which could focus more on early-time cosmological 
signals~\refcite{UV101,SKN,HB01,SP,KT}. The technological challenges for these 
missions are considerable and deserve very careful investigations. 
Finally, ground-based GW detectors in the high-frequency region of MHz~\refcite{EMRussians}
and GHz~\refcite{EM} have gained more attention~\refcite{PBB,G99,KT97}.

So, even if the detection of GWs from primordial Universe is still somewhat 
far ahead of us, maybe, if nature is kind with us, we will not wait 
for a long time. There could be surprises. 

\section*{Acknowledgments}
I wish to thank the organizers of the TASI school for having invited me to 
such a pleasant and stimulating school and
all the students for their interesting questions.
In preparing and writing these lectures I benefited from conversations 
with Yanbei Chen, Scott Hughes, Marc Kamionkowski, Arthur Kosowsky, 
David Langlois, Shane Larson, Albert Lazzarini, 
Michele Maggiore, J\'er\^ome Martin, Alberto Nicolis, Sterl Phinney, Patricia 
Purdue and Kip Thorne. 

This research was also supported by Caltech's Richard Chace Tolman Foundation.

\end{document}